\documentclass[preprint,authoryear,3p,times]{elsarticle}

\usepackage[latin1]{inputenc}

\usepackage{graphicx}
\usepackage{subfigure}
\usepackage{bm}
\usepackage{hyperref}

\usepackage{empheq}
\usepackage{xspace}
\usepackage{color}
\usepackage{amssymb}
\usepackage{booktabs}
\usepackage{appendix}
\usepackage{placeins}
\usepackage[bbgreekl]{mathbbol}

\hypersetup{
bookmarksnumbered = true,
unicode=false,          
pdftoolbar=true,        
pdfmenubar=true,        
pdffitwindow=false,     
pdfstartview={Fit},    
pdfauthor={P.Recho},     
pdfsubject={Subject},   
pdfcreator={Creator},   
pdfproducer={Producer}, 
pdfkeywords={keyword1} {key2} {key3}, 
pdfnewwindow=true,      
colorlinks=true,       
linkcolor=black,          
citecolor=black,        
filecolor=black,      
urlcolor=black,           
pdfdisplaydoctitle = true
}

\usepackage[normalem]{ulem}
\usepackage{color}

\newcommand{\tr}[1]{\textcolor{black}{#1}}

\journal{Journal of the Mechanics and Physics of Solids}

\usepackage[normalem]{ulem} 

\begin{document} 
\text

\title{Mechanical behavior of multi-cellular spheroids under osmotic compression}

\author[liphy,iab]{M. Dolega}
\author[NUI]{G. Zurlo}
\author[liphy]{M. Le Goff}
\author[liphy]{M. Greda}
\author[liphy]{C. Verdier}
\author[college,curie]{J.-F. Joanny}
\author[liphy]{G. Cappello}
\author[liphy]{P. Recho}
\ead{pierre.recho@univ-grenoble-alpes.fr}
\address[liphy]{Universit\'e Grenoble Alpes, Laboratoire Interdisciplinaire de Physique, CNRS, F-38000 Grenoble, France}
\address[iab]{Institute for Advanced Biosciences, Centre de Recherche Universit\'e Grenoble Alpes, Inserm U 1209, CNRS UMR 5309, F-38700 La Tronche, France}
\address[NUI]{School of Mathematics, Statistics and Applied Mathematics, NUI Galway, University Road, Galway, Ireland}
\address[college]{Coll\`ege de France, PSL Research University, 11 place Marcelin Berthelot, 75005 Paris, France}
\address[curie]{Physico-Chimie Curie CNRS--UMR 168, Institut Curie, 11 Rue Pierre et Marie Curie, 75005 Paris, France}

\begin{abstract}
The internal and external  mechanical environment plays an important role in tumorogenesis. As a proxy of an avascular early state tumor, we use multicellular spheroids, a composite material made of cells, extracellular matrix and permeating fluid. We characterize its effective rheology at the timescale of minutes to hours by compressing the aggregates with osmotic shocks and modeling the experimental results with an active poroelastic material that reproduces the stress and strain distributions in the aggregate. The model also  predicts how the emergent bulk modulus of the aggregate as well as the hydraulic diffusion  of the percolating interstitial fluid are \tr{modified by the preexisting active stress within the aggregate}. We further show that the value of these two phenomenological parameters can be rationalized by considering that, \tr{in our experimental context}, the cells are \tr{effectively} impermeable and incompressible inclusions nested in a compressible and permeable matrix.
\end{abstract}

\maketitle

\section{Introduction}

The initial growth phase of a tumor is generally avascular \citep{tracqui2009biophysical}. Three dimensional multi-cellular spheroids (MCS) of living cells  that incorporate cancer cells, extra-cellular matrix (ECM) and a permeating fluid constitute a good model system \citep{lin2008recent} to understand the main physical mechanisms controlling this initial expansion. Among such mechanisms, hindered diffusion of biochemical factors such as nutrients, oxygen or growth factors  that affect the cell proliferation (contact inhibition), differentiation and motility (chemotaxis) play an important role \citep{tracqui2009biophysical}. But it is also known that active and passive mechanical forces stemming from growth induced residual stress and the poro-visco-elasticity of the aggregate and its confining environment control the expansion of a tumor and the creation of new metastasis \citep{tracqui2009biophysical,shieh2011regulation,jain2014role}. A direct proof was brought by the seminal contribution of \cite{helmlinger1997solid} who applied an elastic stress on  MCS by culturing them in porous agarose gels of various stiffnesses to show that the final size reached by the MCS decreases as the stiffness of the gel increases. The same result was later confirmed with a high throughput by \cite{Alessandri2013} who designed a microfluidic device to grow MCS in permeable elastic capsules. Interestingly, a conceptually different protocol has been used by \cite{delarue2013mechanical} to compress MCS by supplementing the culture medium with large Dextran molecules that cannot permeate the spheroid pores. The imposed osmotic pressure creates a gradient of interstitial pore pressure in the MCS and leads to an interstitial fluid flow that dehydrates the spheroid leading to its compression and an elevated pore pressure at the end of the compression. In full agreement with the results of \cite{helmlinger1997solid}, osmotic pressures between 500 and 5000 Pa considerably slow the spheroid growth, mainly by arresting the cell cycle at the end of G1 phase in the spheroid core \citep{delarue2014compressive}. Again, similarly to the results obtained using agarose gel confinement \citep{cheng2009micro,stylianopoulos2012causes}, osmotic compression leads to a solid stress gradient from the periphery to the core of the spheroid \citep{dolega2017cell}. The equivalence of these two protocols further underlines the importance of interstitial flow as a key mechanical component regulating the dynamics of MCS growth \citep{shieh2011regulation}.

The important role played by both solid and fluid mechanics has therefore motivated the rheological characterization of multicellular aggregates (reviewed in \cite{gonzalez2012soft}) as a continnuum medium in order to assess the internal distribution of stresses and strains inside the aggregate and identify the potential mechano-transduction pathways that feedback on the biological mechanisms \citep{tracqui2009biophysical}. As for many soft matter systems, such a rheology is strongly dependent on the characteristic timescale on which the response of the MCS to a mechanical perturbation is monitored \citep{Marmottant2009, preziosi2010elasto}. 

Most of the existing works focus on the long timescale (longer than the typical cell cycle duration, i.e. several hours at least) where the fundamental question is how cell growth coordinates with the mechanical stress inside the MCS. Single phase continuum models using the framework of morphoelasticity \citep{goriely2017mathematics} embed growth and remodelling as an incompatible pre-strain in the framework of  non-linear  elasticity \citep{Ambrosi2004, Ciarletta2013}. Following a close-to-equilibrium thermodynamics Onsager relation, the pre-strain itself is driven by a generalized Eshelby driving force towards a homeostatic state that depends on the mechanical stress and nutrient/oxygen concentrations  \citep{Ambrosi2007,Ambrosi2011,ambrosi2017solid}. If cell division and apoptosis are fast enough, cell rearrangements effectively result in an active viscosity and pressure \citep{Ambrosi2008,Ranft2010,delarue2014stress} justifying the use of fluid models. Multiphase mixture theories minimally involving  a cellular and the extra-cellular phase have also been used with the idea that cell proliferation and/or matrix production and degradation can be accounted for by mass exchange terms between the phases \citep{Humphrey2002,Byrne2003,Roose2003,Mascheroni2016}. These models can be used in tandem with the morphoelastic framework \citep{Ambrosi2008,giverso2012modelling}. Recently, the classical framework of poroelasticity  \citep{biot1941general,rice1976some, coussy2004poromechanics} has been augmented to account for cellular turnover using again the morphoelastic framework where the growth tensor is controlled in a thermodynamically consistent way by diffusive solute molecules \citep{xue2016biochemomechanical}. The advantage of such a model is that it also accounts for the hydraulic motion of the signaling molecules in the interstitial space. A similar framework has been enriched to account for the competition between healthy and cancer cells during  cancer growth \citep{fraldi2018cells}. On the other side of the spectrum, short timescales (shorter than the typical time needed for water to percolate in the MCS, i.e. few seconds) have also been recently investigated using a poroelastic model \citep{Margueritat2019}.

In this paper, we focus on the intermediate timescale (typically few tens of minutes) where the interstitial fluid has time to permeate in and out the spheroid but where the cell proliferation is negligible so that the cells do not significantly change neighbors in the course of the experiments. Our aim is to quantitatively characterize the MCS rheology at that time scale using osmotic compression experiments. Note that the active stress induced by growth is still present in this situation, but is essentially frozen at the timescale of the mechanical test. For this reason, we use a poroelastic framework where a fixed but inhomogeneous  active stress is embedded in the material. In particular, we predict and  check experimentally how the effective bulk modulus and interstitial water hydrodynamic diffusion are renormalized by the cellular activity during the compression test, which extends the ground-breaking work of \cite{netti2000role} obtained by direct mechanical compression of a MCS in a confining chamber. Using experimental characterization of both cells and an ECM proxy, we also show how the effective MCS rheological properties can be rationalized by considering that cells are impermeable and incompressible inclusions in a soft poroelastic ECM permeated by interstitial water. Electrostatic effects due to fixed charges carried by some ECM components \citep{Mow1998,Xue2017} are not accounted for in the present model. Experimentally, \cite{voutouri2014evolution} pointed out the role of the ECM electro-osmotic swelling in tumor growth. Because the ECM of tumors contains negatively charged macro-molecules, counter-ions permeate the MCS to establish electroneutrality and create an osmotic pressure difference which is equilibrated by an interstitial hydrostatic pore pressure within the MCS that swells the ECM. However high  modifications of counter ions concentrations (200 mM NaCl) are necessary to record a non-negligible pore pressure increase \citep{voutouri2016hyaluronan}. 

The paper is organized as follows. In Section~\ref{sec:model_formulation} we establish the physical equilibrium laws and discuss the assumptions related to our active poroelastic model based on an active stress which is inhomogeneous and time independent. The conditions satisfied by these slow mechanical variables are then given in Section~\ref{sec:before_shock}.  Section~\ref{sec:passive} briefly recalls the classical results that can be derived from our framework in the passive case. In Section~\ref{sec:active}, we specify the ansatz for the active stress in a way that derives from previous experimental and theoretical works. Based on this ansatz, we compute the resulting stress and strain and use the experimental data to fit our model parameters in Section~\ref{sec:stress_strain}. Knowing the strain distribution, we find the ensuing effective bulk modulus and hydraulic mobility (Section~\ref{sec:water_relaxation}) of the MCS which we compare, with no adjustable parameters, to our experimental results. In Section~\ref{sec:composite}, we rationalize the values of the effective bulk modulus and hydraulic mobility by showing how they can emerge from a situation where the MCS is considered as a composite material with incompressible and impermeable cells (at the timescale of our experiment) nested in a poroelastic matrix represented by the ECM. Finally, we summarize our results in Section~\ref{sec:conclusion}. Several appendices gather auxiliary results and  experimental procedures.

\section{Model formulation}\label{sec:model_formulation}

Our model aims at capturing the deformation of a MCS in response to an osmotic shock (i.e. an abrupt increase of the osmolarity of the medium in which the MCS is cultured). Since the deformation of the MCS following the shock occurs over a timescale of a few minutes, we model the spheroid as an poroelastic active  material. 

\paragraph{Kinematics} Before the application of the osmotic shock, the MCS occupies the domain $\Omega$, a ball of center $0$ and radius $R_m$.  We denote the space coordinate $\underline{X}\in\Omega$ and the time $t\geq 0$. The boundary of $\Omega$ is denoted by $\partial\Omega$ and the outward normal at a point  $\underline{X}$ of the boundary is $\underline{N}$. Because of the external osmotic loading, the MCS is deformed from its initial configuration $\Omega$ at time $t=0$ to a new configuration $\omega$ at time $t$. Subsequently, material points at position $\underline{x}\in \omega$ are mapped from the initial configuration to the actual one by the transformation $\underline{x}=\phi(\underline{X},t)$ whose gradient is the deformation tensor $\mathbb{F}(\underline{X},t)=\nabla_X \phi$. Built on $\mathbb{F}$, we consider the right Cauchy-Green strain measure: $\mathbb{E}=(\mathbb{F}^T\mathbb{F}-\mathbb{I})/2$ where $\mathbb{I}$ is the identity. When the displacement $\underline{u}(\underline{X},t)$ from the initial configuration is small (i.e. $\nabla\underline{u}\ll 1$), $\mathbb{E}=(\nabla\underline{u}^T+\nabla\underline{u})/2$ reduces to the usual linear strain that we will use in the model presented below.

\paragraph{Momentum balance}
In the absence of inertia and external forces, The first Piola-Kirchhoff stress tensor $\mathbb{P}(\underline{X},t)$ satisfies the force balance equation,
\begin{equation}\label{e:force_bal}
\nabla.\mathbb{P}=0 \text{ with boundary condition (B.C.) } \mathbb{P}\vert_{\partial\Omega}\underline{N}=0.
\end{equation}
We measure the stress with respect to the hydrostatic pressure in the external fluid, which implies the absence of traction force at the MCS boundary. The balance of torques imposes the symmetry condition on the stress $\mathbb{F}\mathbb{P}^T=\mathbb{P}\mathbb{F}^T$. Note that $\mathbb{P}$ is the total stress encompassing both the solid and fluid contributions \citep{jain2014role} in the MCS.

\paragraph{Mass balance} Assuming that the internal flow of extra-cellular fluid follows a Darcy law, mass conservation of the incompressible fluid (mass density $\rho$) can be expressed in  the following way \tr{(see \ref{sec:thermo_found} for details)}:
\begin{equation}\label{e:const_be}
\frac{1}{\rho}\frac{\partial m}{\partial t}-\frac{\kappa}{\mu}\nabla^2p=\frac{S}{\rho},
\end{equation}
where $m(\underline{X},t)$, homogeneous to a mass density, is the extra-cellular fluid mass per reference volume element,  \tr{$\kappa$ is the assumed isotropic MCS effective permeability (m$^2$) before the application of the osmotic shock}, $\mu$ (Pa.s) is the extra-cellular fluid viscosity, $p(\underline{X},t)$ the interstitial pressure in the MCS intercellular pores and $S(\underline{X})$ is a source term \citep{Roose2003, fraldi2018cells}  representing the fact that cells die in the center of the spheroid, therefore producing interstitial fluid, while cells divide and grow at the periphery, therefore uptaking interstitial water \citep{delarue2013mechanical}. This rationale relies on the fact that cells growth is tightly linked to fluid exchange since cells are mostly made of fluid \citep{cadart2019physics}. Given our timescale of interest, we consider that $S$ is time independent.

To derive the boundary conditions associated to \eqref{e:const_be}, we conceptually imagine the presence of a very compliant dialysis bag around the MCS, which lets water and ions go through but is impermeable to large macromolecules used to perform the osmotic shock. If the filtration coefficient of the bag is denoted by $L_p$, the water flux through the bag is  $L_p(\triangle \Pi-\triangle p)$ where  $\Pi$ and $p$ are the osmotic and hydrostatic pressures and $\triangle$ denotes the difference between the two sides of the bag \citep{kedem1958thermodynamic}. Taking the limit $L_p$ large (infinite permeability of the bag), we obtain $\triangle \Pi=\triangle p$. Choosing the pressure outside the spheroid as $p=0$ and the denoting the external imposed osmotic pressure as $\Pi_e(t)$  leads to 
$$p\vert_{\partial\Omega}=\delta \Pi_a-\Pi_e,$$
where $\delta \Pi_a$ is a difference of osmotic pressure between the spheroid and the external culture medium, which is actively maintained by cellular ion pumps \citep{cadart2019physics}. \tr{Importantly, to perform the osmotic shock, we use dextran molecules with a radius of gyration larger than 15 nm that are too large to permeate through the MCS pores (see \ref{sec:poro_ECM}) and are therefore globally excluded from the MCS.}

\paragraph{Constitutive behavior}
We introduce the mechanical energy of the spheroid $g$, to express the interstitial pressure and the Piola-Kirchhoff stress as \tr{(see \ref{sec:thermo_found} for details)}
\begin{equation}\label{e:thermo_rel}
p=\rho\frac{\partial g}{\partial m} \text{ and } \mathbb{P}=\mathbb{F}\frac{\partial g}{\partial \mathbb{E}}. 
\end{equation}
\tr{As we approximate (see again \ref{sec:thermo_found}) $g$ by its expansion up to quadratic order it reads
$$g=g_0+\mathbb{P}_a:\mathbb{E}+\frac{p_a}{\rho}\delta m+\frac{K_u}{2}\text{tr}(\mathbb{E})^2+G \mathbb{E}:\mathbb{E}-\zeta \text{tr}(\mathbb{E})\delta m+\frac{\chi}{2}\delta m^2,$$
in which the active stress $\mathbb{P}_a(\underline{X})$ and interstitial pressure $p_a(\underline{X})$ are time  independent and $\delta m$ is the variation of fluid mass per unit element volume compared to the initial configuration.} In order to account for the active stress in the material response of the spheroid in an analytically tractable way, we assume that the displacement $\underline{u}$ is small, so that, \tr{truncating formulas \eqref{e:thermo_rel} to linear order in $\underline{u}$,  the current stress and interstitial pressure are given by 
\begin{equation}\label{e:free_energ_var}
\begin{array}{c}
\mathbb{P}=\mathbb{P}_a+(\nabla\underline{u})\mathbb{P}_a+\mathbb{L}[\mathbb{E}]-b(p-p_a)\mathbb{I}\\
\frac{\delta m}{\rho}=\frac{b^2}{K_u-K_d}(p-p_a)+b \text{tr}(\mathbb{E}),
\end{array}
\end{equation}
where the so-called tangential operator \citep{paroni2009variational} $\mathbb{L}[\mathbb{E}]$ assumes a Saint-Venant Kirchhoff form
$$\mathbb{L}[\mathbb{E}]=K_d\text{tr}(\mathbb{E})\mathbb{I}+2G\mathbb{E}.
$$
In \eqref{e:free_energ_var}, we have introduced four conventional \citep{coussy2004poromechanics} constitutive parameters. $K_u$ and $K_d$  are the undrained and drained bulk moduli of the spheroid corresponding respectively to a situation where the water cannot flow out during deformation and where the water is able to flow out. $G$ is the shear modulus and  $b$ is the (dimensionless) Biot coefficient. The coefficients $\zeta$ and $\chi$ of the mechanical energy can be related to these classical parameters through the relations:
$$K_u-K_d=\frac{\zeta^2}{\chi} \text{ and } b=\frac{\zeta}{\rho\chi}.$$
We note that in general $K_u$, $K_d$, $G$ and $b$  should depend on the underlying state of  active stress and deformation (see \ref{sec:thermo_found}), but for simplicity we assume here that they are constant.}

\paragraph{Simplifications} We assume, as for most biological tissues \citep{Cowin2007,fraldi2018cells}, that $b\simeq 1$  and $K_u\gg K_d$, as it is very difficult to compress the MCS in an undrained situation.  Thus we are left with only three rheological parameters, $K_d$, $G$, $\kappa/\mu$ determining the response of the MCS to the osmotic compression. The final problem therefore reads

\begin{equation}\label{e:problem_poro_elast}
\begin{array}{c}
\nabla.\mathbb{P}=0 \text{ with B. C. } \mathbb{P}\vert_{\partial\Omega}\underline{N}=0.\\
\frac{\partial \delta m}{\partial t}-\rho\frac{\kappa}{\mu}\nabla^2p=S,\text{  with B. C. }p\vert_{\partial\Omega}=\delta \Pi_a-\Pi_e\\
\mathbb{P}=\mathbb{P}_a+{\color{black}(\nabla\underline{u})}\mathbb{P}_a+\mathbb{L}[\mathbb{E}]-(p-p_a)\mathbb{I}\text{ and }
\delta m=\rho\text{tr}(\mathbb{E}).
\end{array}
\end{equation}

\section{Equilibrium before the osmotic perturbation}\label{sec:before_shock}

Before any osmotic stress is applied ($\Pi_e=0$), $\mathbb{E}=0$ and $\delta m=0$ such that $\mathbb{P}=\mathbb{P}_a$ and $p=p_a$. Problem \eqref{e:problem_poro_elast} naturally requires that forces are equilibrated and mass is balanced in this initial state such that the fields $\mathbb{P}_a$ and $p_a$ satisfy:  
\begin{equation}\label{e:conditions_Pa}
\nabla.\mathbb{P}_a=0 \text{ with boundary condition } \mathbb{P}_a\vert_{\partial\Omega}\underline{N}=0 \text{ and }\mathbb{P}_a=\mathbb{P}_a^T.
\end{equation}
and 
\begin{equation}\label{e:conditions_pa}
-\frac{\kappa}{\mu}\nabla^2p_a=\frac{S}{\rho} \text{ with boundary condition }p_a\vert_{\partial\Omega}=\delta \Pi_a 
\end{equation}
Injecting these relations back into \eqref{e:problem_poro_elast} and defining $\delta p=p-p_a$, we obtain the active poroelastic problem ruling the MCS deformation and interstitial pressure upon osmotic compression:
\begin{equation}\label{e:problem_poro_elast_simpl}
\begin{array}{c}
\nabla.({\color{black}(\nabla\underline{u})}\mathbb{P}_a+\mathbb{L}[\mathbb{E}])=\nabla \delta p \text{ with B. C. } ({\color{black}(\nabla\underline{u})}\mathbb{P}_a+\mathbb{L}[\mathbb{E}])\vert_{\partial\Omega}.\underline{N}= \delta p\vert_{\partial\Omega}\underline{N}\\
\frac{\partial \text{tr}(\mathbb{E})}{\partial t}-\frac{\kappa}{\mu}\nabla^2\delta p=0,\text{  with B. C. }\delta p\vert_{\partial\Omega}=-\Pi_e.
\end{array}
\end{equation}
The specificity of this problem is that it depends on the active stress tensor $\mathbb{P}_a$.

\section{The passive response}\label{sec:passive}

In the absence of cellular activity, $\mathbb{P}_a=0$ and \eqref{e:problem_poro_elast_simpl} reduces to a classical linear poroelastic problem. In this case, taking the divergence of the mechanical equilibrium, we obtain, $\nabla^2p=(4G/3+K_d)\nabla^2 \text{tr}(\mathbb{E})$ and \eqref{e:problem_poro_elast} becomes
\begin{equation}\label{e:problem_poro_elast_lin}
\begin{array}{c}
\nabla.(\mathbb{L}(\mathbb{E}))=\nabla \delta p \text{ with B. C. } \mathbb{L}(\mathbb{E})\vert_{\partial\Omega}\underline{N}=\delta p\vert_{\partial\Omega}\underline{N} \\
\frac{\partial\text{tr}(\mathbb{E})}{\partial t}-\frac{\kappa(4G/3+K_d)}{\mu}\nabla^2\text{tr}(\mathbb{E})=0\text{  with B. C. }\delta p\vert_{\partial\Omega}=-\Pi_e.
\end{array}
\end{equation}
From \eqref{e:problem_poro_elast_lin}  we deduce two classical results:
\begin{enumerate}
\item After the osmotic shock, the water percolates out of the MCS following a diffusion process with a hydraulic diffusion coefficient \citep{biot1941general,tanaka1979kinetics} 
$$D_{\text{passive}}=\kappa(4G/3+K_d)/\mu,$$
until a steady state is reached where the interstitial pore pressure equates minus the imposed osmotic pressure throughout the whole MCS.
\item In this final state, the volumetric strain within the MCS is constant $\text{tr}(\mathbb{E})=-\Pi_e/K_d$. Hence, the relative change of volume of the MCS following the shock is associated to the drained compressibility modulus (or osmotic modulus) 
$$K_{\text{passive}}=K_d.$$
\end{enumerate}
The aim of this paper is to generalize these two results to the active case in a spherical geometry and to compare the predictions with experiments.

\section{The active response}\label{sec:active}

In the rest of the paper, we assume that problem \eqref{e:problem_poro_elast_simpl} has a spherical symmetry such that all the considered fields only depend on the spherical coordinate $\underline{X}=R\underline{e}_R$ where $\underline{e}_R$ is the radial basis vector. The radial symmetry implies that the displacement field reduces to an unknown scalar $\underline{u}=u(R)\underline{e}_R$ and 
$$\mathbb{E}=\left( \begin{array}{ccc}\partial_Ru(R)&0&0\\
0&u(R)/R&0\\
0&0&u(R)/R
\end{array}\right)\text{ and }\mathbb{P}_a=\left( \begin{array}{ccc}P^a_r(R)&0&0\\
0&P^a_{\theta}(R)&0\\
0&0&P^a_{\theta}(R)
\end{array}\right),$$
From the initial mechanical equilibrium \eqref{e:conditions_Pa}, we obtain that the components of $\mathbb{P}_a$ satisfy 
\begin{equation}\label{eq:hoop_pre_stress}
P^a_{\theta}=\frac{1}{2}\left(2P^a_r+R\partial_RP^a_r \right)
\end{equation}
and plugging this expression into the mechanical equilibrium relation in \eqref{e:problem_poro_elast_simpl}, we obtain the differential equation on $u$ 
\begin{equation}
R^2 \partial_{RR}u (4 G+3 K_d+3 P^a_r)-\left(u-R \partial_Ru\right) \left(8 G+6 K_d+3 R \partial_RP^a_r+6 P^a_r\right)-3 R^2 \partial_Rp=0.
\end{equation}
By classical methods, we obtain the general solution of such a problem:
$$u(R,t)=c_1 R f(R)+c_2 R+3 R \int_R^{R_m} v^3 (f(R)-f(v)) \partial_vp(v,t) \, dv,$$
where $c_1$ and $c_2$ are two integration constants and the function $f$ reads
$$f(R)=\int_R^{R_m} \frac{dq}{q^4 (4 G+3 K_d+3 P^a_r(q))}  .$$
Based on the long time scale theory presented in \cite{delarue2014stress}, we consider the following special form of $P^a_r(R)$ satisfying the boundary conditions of \eqref{e:conditions_Pa} :
\begin{equation}\label{eq:pre_stress_compo}
P^a_r(R)=P_a \left(1-\left(\frac{R_m}{R}\right)^{\beta}\right) \text{ leading through \eqref{eq:hoop_pre_stress} to, } P^a_{\theta}(R)=P_a \left(\left(\frac{\beta }{2}-1\right) \left(\frac{R_m}{R}\right)^{\beta }+1\right),
\end{equation}
where $P_a$ is the magnitude of the active stress and $\beta$ an exponent characterizing the spatial variation of the internal active stress. The domains where these two new parameters can vary will be investigated in the following section.

\section{Steady state stress and strain in the MCS}\label{sec:stress_strain} 
\begin{figure}[h!]
\centering
\includegraphics[width=1.0\textwidth]{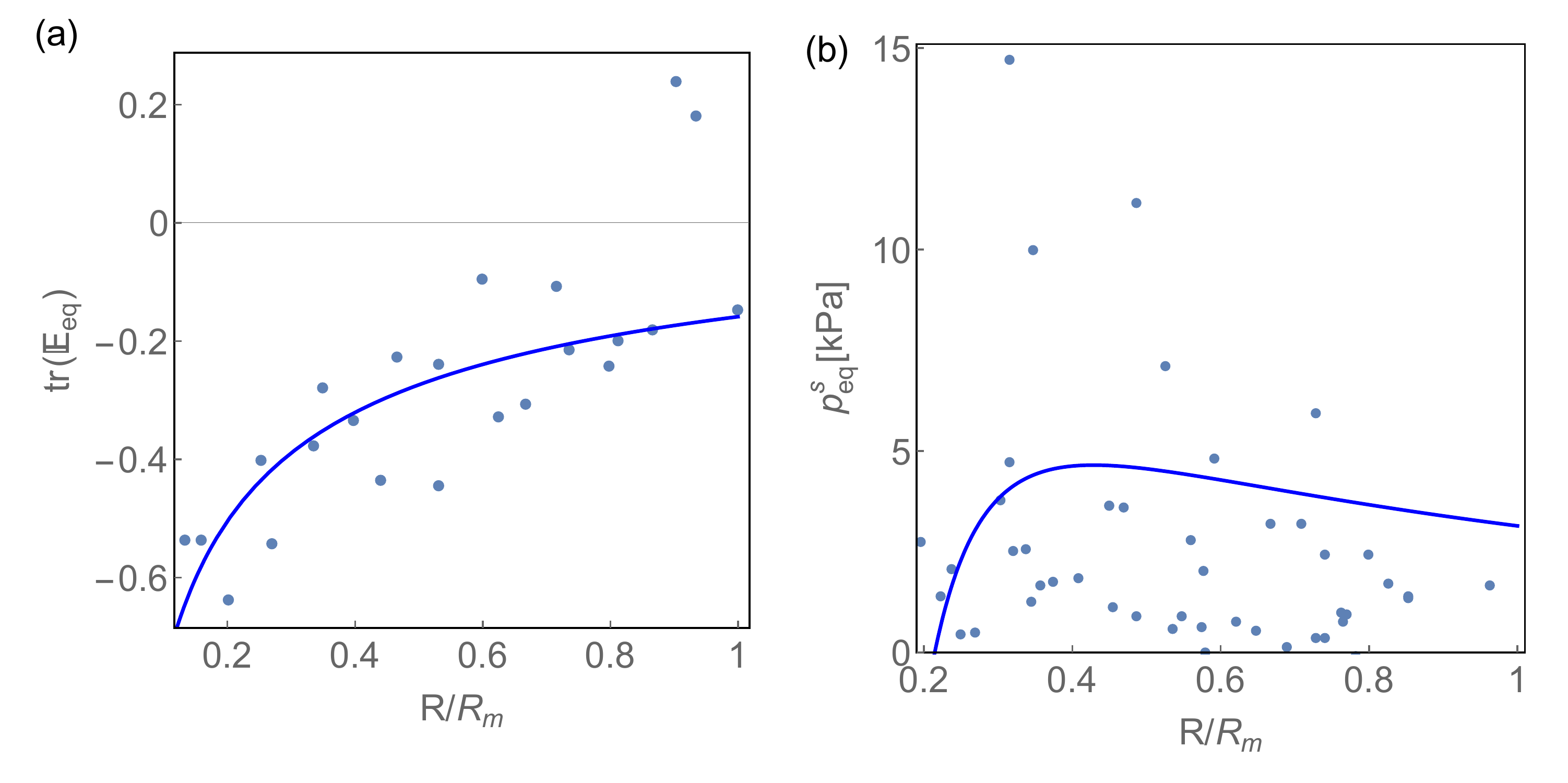}
\caption{\label{fig:Fitstress} \textbf{\tr{Non-deviatoric part of the} strain and stress in the MCS}. Full lines are the theoretical predictions, blue dots correspond to the raw experimental data (a) Steady state volumetric strain under a $\Pi_e=5$kPa osmotic compression estimated from the variation of the nucleus distance (data from \cite{delarue2014stress}). The fitted parameters are $K_d\simeq P_a\simeq 30$ kPa and $\alpha\simeq -0.8$. (b) Steady state pressure $p^s_{\text{eq}}$ within the MCS under $\Pi_e=5$kPa osmotic compression  (data from \cite{dolega2017cell}). The full line corresponds to the analytic expression $p^s_{\text{eq}}$ with parameters from (a). The polyacrylamide beads that serve as local stress gauges in \cite{dolega2017cell} are internalized within the MCS before the osmotic stress is applied. Beads are permeable to water so that the pore pressure equilibrates between the intercellular space and the beads. They are therefore sensitive to the pressure field $p^s_{\text{eq}}$.  }
\end{figure}

Following the osmotic compression, the MCS loses volume and reaches a new steady state which is time-independent, at a time scale much smaller than the typical duration of cell division, which dictates the temporal evolution of the active stress $\mathbb{P}_a$ and interstitial pressure $p_a$. In this situation, $\delta p=-\Pi_e$ is a constant and 
$$u_{\text{eq}}(R)=c_1 R f(R)+c_2 R.$$
To find the integration constants, we first impose the zero stress boundary condition of \eqref{e:problem_poro_elast_simpl} at the interface between the spheroid and the culture medium. The second condition is that $u_{\text{eq}}(0)=0$ by symmetry of the problem.  But as soon as $\beta \leq 2$, we have that $\lim _{R\rightarrow 0}Rf(R)\neq 0$, which implies that  $c_1=0$ and $u_{\text{eq}}$ is linear in $R$ as in a passive problem. This entails a constant volumetric strain $\text{tr}(\mathbb{E}_{\text{eq}})$, not in agreement with the experimental results (See Fig.~\ref{fig:Fitstress}~(a)). We thus conclude that we necessarily have $\beta > 2$. The upper bound for $\beta<3$ appears from the experimental observation that $\text{tr}(\mathbb{E}_{\text{eq}})$ diverges  for small values of $R$. Additionally, this condition is also necessary for the total active stress in the MCS to be finite (i.e. $R^2P^a_r(R)$ integrable in the vicinity of $R=0$). We therefore consider that $\beta$ lies in this range:
$$2<\beta<3.$$ 
For such values of $\beta$, it is clear from \eqref{eq:pre_stress_compo} that we need to have $P_a<0$ in order that the hoop active stress  $P^a_{\theta}$ is negative in the MCS core. This sign of the hoop stress is suggested by cutting experiments (See \cite{stylianopoulos2012causes} for the seminal experiments on real tumors and \cite{colin2018experimental, guillaume2019characterization} for MCS similar to the ones used in this paper). However, in these experiments, the measurements concern the solid stress in the MCS, which is \emph{not} $\mathbb{P}_a$ but a combination of $\mathbb{P}_a$ and $p_a$. Denoting $n_0$ the initial porosity of the MCS and assuming a linear superposition of the initial solid and fluid stress contributions in the MCS, we have $\mathbb{P}_a=(1-n_0)\mathbb{P}_a^s-n_0p_a\mathbb{I}$ where $\mathbb{P}_a^s$ is the solid stress inferred from the deformation of the MCS following a cut in the radial direction. In the aforementioned experiments, the hoop stress related to $\mathbb{P}_a^s$ is positive at the MCS surface and negative in the core. This is consistent with our ansatz of $\mathbb{P}_a$ if $p_a$ remains bounded in the core (as shown in \ref{sec:pa}) since the sign of the hoop solid stress in the core and at the periphery are respectively given by:
$$P^a_{\theta}+n_0p_a\underset{R=0}{\sim}\frac{P_a}{2} (\beta -2)  \left(\frac{R_m}{R}\right)^{\beta }<0 \text{ and }P^a_{\theta}+n_0p_a\underset{R=R_m}{\sim}\frac{\beta P_a}{2}+n_0\delta \Pi_a>0.$$
The first inequality follows from our restriction that $P_a<0$ and the second one requires that $\delta \Pi_a$ is sufficiently large.

With these restrictions on the parameters $\beta$ and $P_a$ characterizing the active stress, the displacement $u_{\text{eq}}$ reads,
\begin{equation}\label{e:steady_state_displ}
u_{\text{eq}}(R)=-\Pi_e\left( \frac{\alpha  R_m^3 \, \mathcal{F}\left(1,\frac{3}{\beta };\frac{\beta +3}{\beta };\frac{3 P_a \left(\frac{R_m}{R}\right)^{\beta }}{4 G+3 (K_d+P_a)}\right)}{R^2 \left(-3 K_d \,
   \mathcal{F}\left(1,\frac{3}{\beta };\frac{\beta +3}{\beta };\frac{3 P_a}{4 G+3 (K_d+P_a)}\right)+4 G+3 (K_d+P_a)\right)}+\frac{(\alpha +1)  R}{3 K_d}\right) 
\end{equation}  
where $\mathcal{F}$ is the ordinary hypergeometric function defined by the power series:
$$\mathcal{F}(a,b;c;z) = \sum_{n=0}^\infty \frac{(a)_n (b)_n}{(c)_n} \frac{z^n}{n!} \text{ with } (q)_n = \begin{cases}  1  & n = 0 \\
  q(q+1) \cdots (q+n-1) & n > 0
 \end{cases}$$
Note that, on top of the already introduced rheological parameters characterizing the active and passive mechanical behaviors of the MCS, expression \eqref{e:steady_state_displ} also contains a free non-dimensional parameter $\alpha$ since the symmetry condition $u_{\text{eq}}(0)=0$ is automatically satisfied when $\beta>2$. Such a parameter $\alpha$ is also of active origin (i.e. there exists a relation $\alpha(P_a)$ such that $\alpha(0)=0$). Indeed when $P_a=0$, the displacement field should reduce to the classical one $u_{\text{eq}}(R)=-\Pi_e R/(3 K_d)$. However, it is yet unclear how to obtain such a relation, for instance via minimization of a certain energy among the family of solutions parametrized by $\alpha$, like in a classical buckling problem.

 With the displacement field \eqref{e:steady_state_displ}, we can compute the volumetric strain $\text{tr}(\mathbb{E}_{\text{eq}})$ and the solid pressure due to the osmotically induced deformation $p^s_{\text{eq}}=-\text{tr}({\color{black}(\nabla\underline{u}_{\text{eq}})}\mathbb{P}_a+\mathbb{L}[\mathbb{E}_{\text{eq}}])/3$, which are both measured experimentally. 

\begin{figure}[h!]
\centering
\includegraphics[width=.7\textwidth]{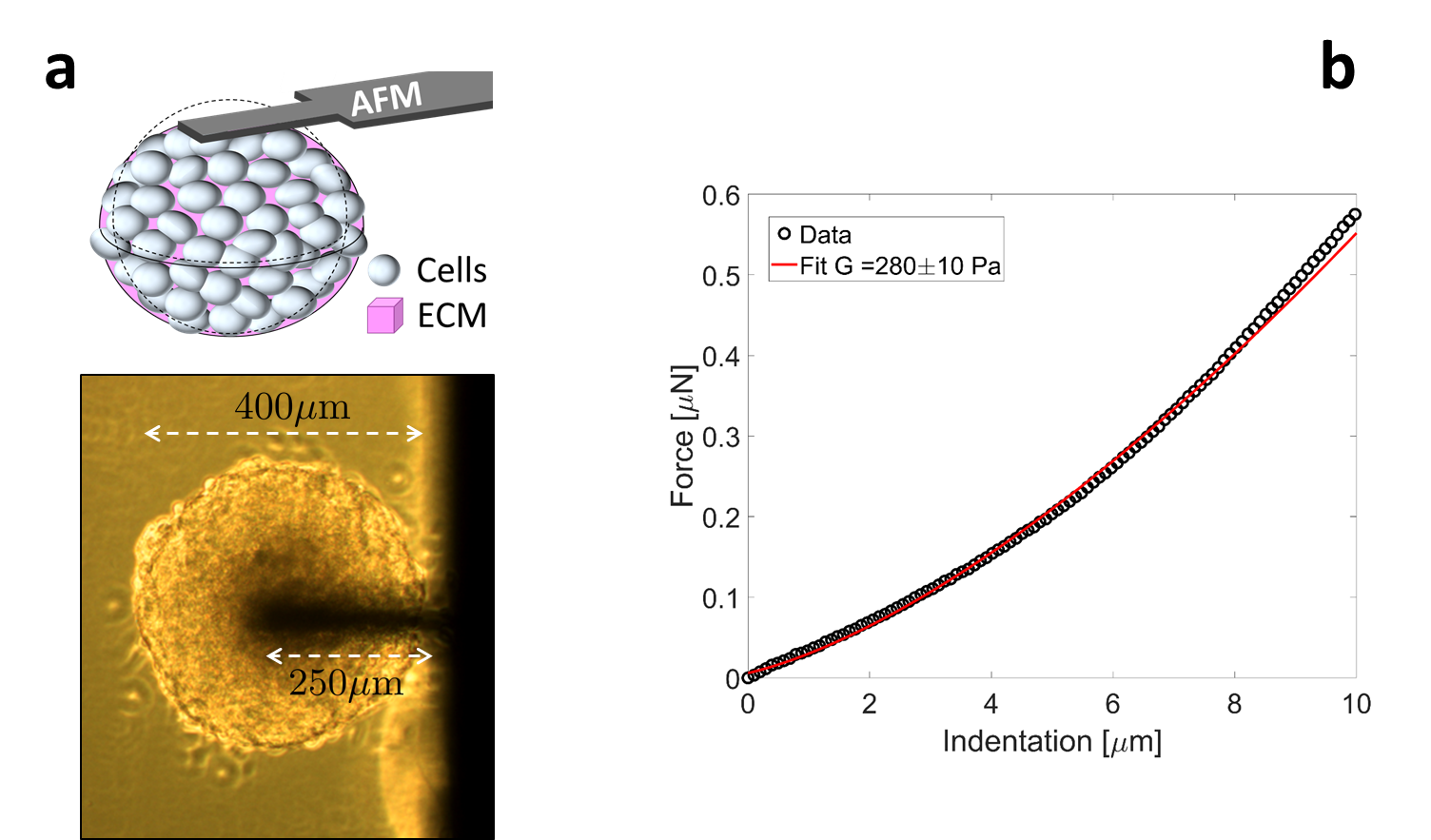}
\caption{\label{fig:MCS_Shear} \tr{\textbf{AFM compression of a spheroid.} (a) Sketch and image of a MCS in contact with the AFM cantilever. (b) Fast ($<500$ ms) force response of a MCS indented at constant speed (20 $\mu$m/s) and fitted using Hertz's model. The best fit is obtained for a shear modulus $E^u =  840\pm20$ Pa.}}
\end{figure}

We show in Fig.~\ref{fig:Fitstress} the curves predicted by the model superimposed with the raw experimental data. To obtain these curves, we arbitrarily fixed $\beta=5/2$ in the middle of the admissible interval and \tr{we measured (experimental protocol described in \ref{Appendix:shear_MCS}) the shear modulus $G$ using an identation AFM setup (see Fig.~\ref{fig:MCS_Shear}~a). We analyzed the initial part of the Force-Indentation curve at a  timescale $< 500$ ms, much shorter than the one of water relaxation where the MCS can be considered as undrained. At small indentation ($<10\mu$m), the curve fits well with Hertz's model (Fig.~\ref{fig:MCS_Shear}~b) for a value of the  undrained Young modulus $E^u =  840\pm20$ Pa and, consequently, $G\simeq E^u/3 = 280\pm10$  Pa (using an infinite value for the undrained bulk mudulus $K_u$). This value is consistent with that found by \cite{Guevorkian2010} on similar MCS.}
We then fit the remaining parameters $K_d$, $P_a$ and $\alpha$ using the \tr{volumetric strain measurements presented in Fig.~\ref{fig:Fitstress}~(a).} Next, we use these parameters to obtain the \tr{non deviatoric part of the solid stress $p^s_{\text{eq}}$ due to the osmotic compression} which we superimpose on the measurements in Fig.~\ref{fig:Fitstress}~(b). Hence, no fitting is done at that level which explains why the agreement is only qualitative. \tr{Also note the large variability in the experimental results which can be partly attributed to the difficulty to find the equatorial plane of the beads serving as stress gauge and hence precisely find their degree of compression after the shock.} 

Interestingly, our expression of $p^s_{\text{eq}}$ gives a plausible explanation for the ``pressure jump'' reported in \cite{dolega2017cell} between the value of $p^s_{\text{eq}}(R_m)$ at the MCS surface and the osmotic pressure $\Pi_e$. Note that $\Pi_e-p^s_{\text{eq}}(R_m)\geq 0$ vanishes in the absence of active stress ($P_a=0$) showing that this jump is of active origin.

\begin{figure}[h!]
\centering
\includegraphics[width=1.0\textwidth]{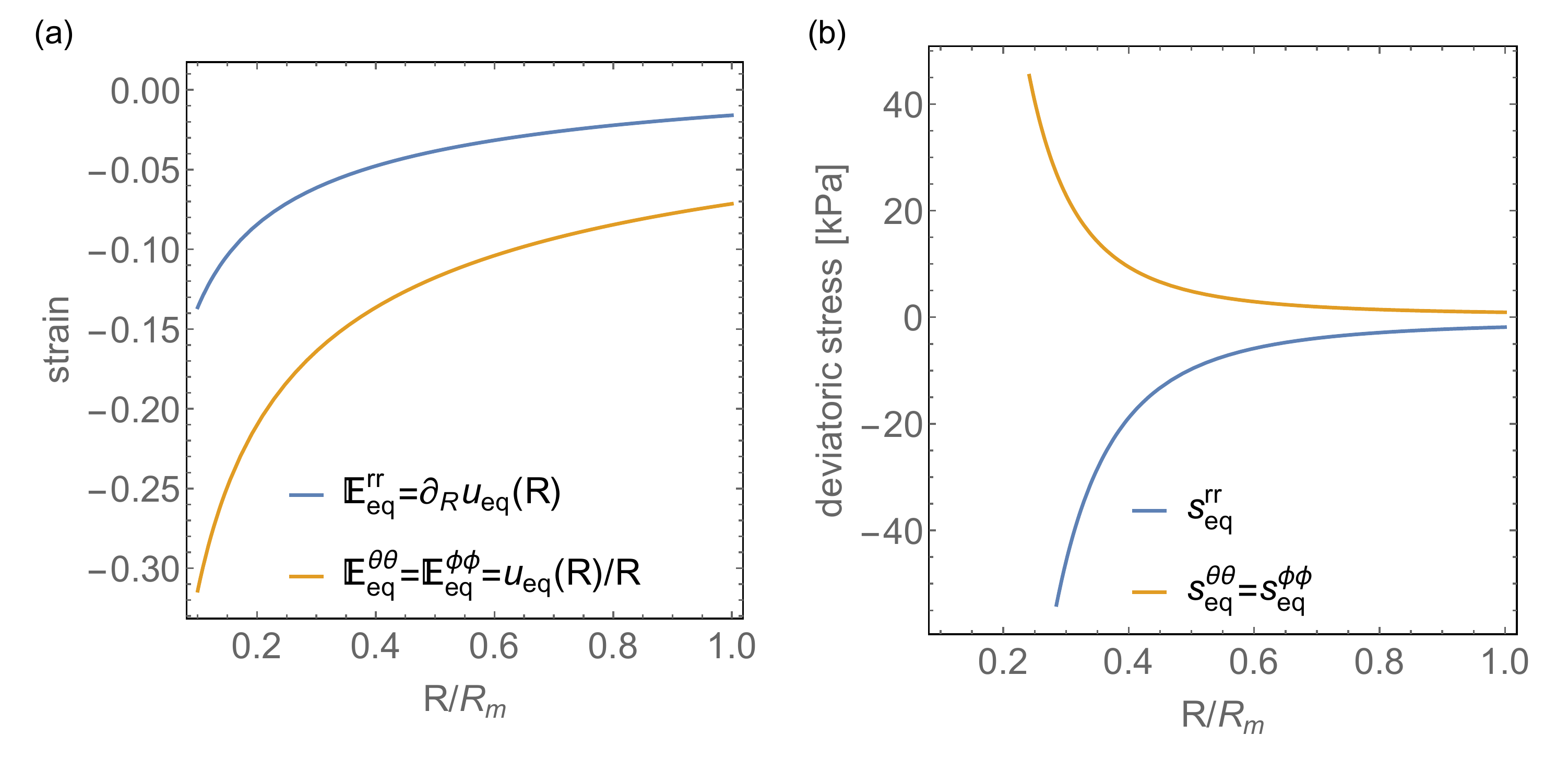}
\caption{\label{fig:anisotropy} \tr{\textbf{ Strain and stress anisotropy due to the compression in the MCS}. (a) Theoretical steady state strain radial and tangential components under a $\Pi_e=5$kPa osmotic compression  (b)Theoretical steady state deviatoric stress radial and tangential components under a $\Pi_e=5$kPa osmotic compression.}  }
\end{figure}

\tr{We also show in Fig.~\ref{fig:anisotropy}, the radial and tangential components of the strain field and the deviatoric part of the steady state solid stress due to the osmotic compression, $s_{\text{eq}}=(\nabla\underline{u}_{\text{eq}})\mathbb{P}_a+\mathbb{L}[\mathbb{E}_{\text{eq}}]+p^s_{\text{eq}}$. Both components of the strain are negative indicating an overall compression of the cells due to the shock but the hoop component is larger. The hoop component of the deviatoric stress is positive while its radial counterpart is negative, therefore cells are compressed in the radial direction and under tension in the tangential direction. One should however not directly associate this anisotropy with the anisotropy distribution of the shape of cells prior to the osmotic compression reported in \cite{dolega2017cell} since this last property is established over a long timescale and is therefore rather reflected in the anisotropy of the stress field $\mathbb{P}_a^s$.}

\begin{figure}[h!]
\centering
\includegraphics[width=1.0\textwidth]{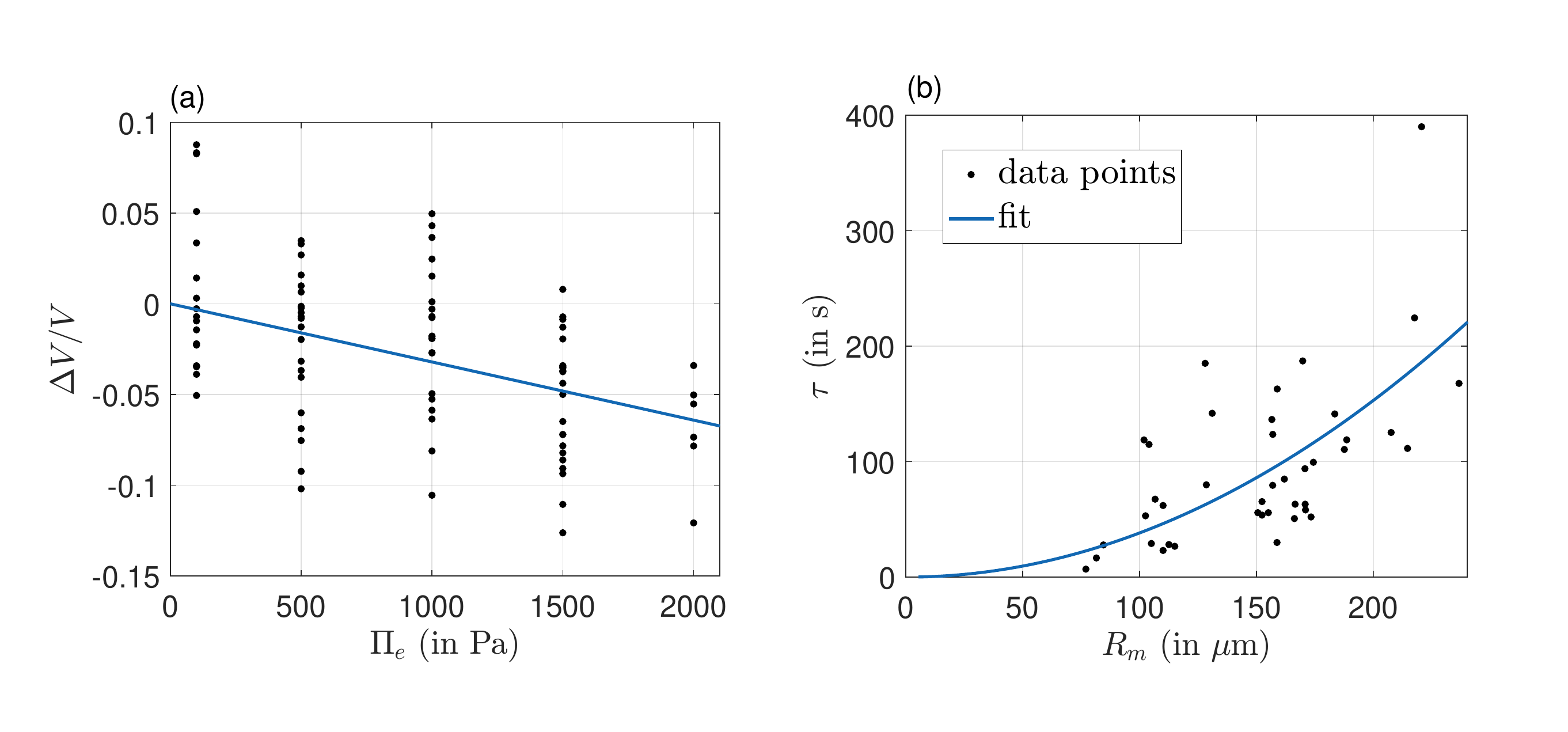}
\caption{\label{fig:compression_diffusion} \textbf{Global stiffness and relaxation of the MCS.} (a) Total volume loss measured when MCS are submitted to a small external osmotic pressure. Dots are experimental data points and the full line is a linear fit with slope $K_{\text{exp}}\simeq 28\in (21,34)$ kPa. Experimentally, the volume is deduced by measuring the projected area of MCS, imaged by phase contrast respectively 5 minutes before and 20 minutes after the osmotic shock. This timescale is short as compared to the cell division time (18 h for CT26 cells).   (b)~Relaxation time of the MCS to its steady radius after an osmotic shock as a function of the MCS initial radius. The fit corresponds to a diffusive ansatz $\tau=R_m^2/(6D_{\text{exp}})$ with a fitted diffusion coefficient $D_{\text{exp}}\simeq 4.4\in(4,4.8)\times 10^{-11}\mu\text{m}^2\text{s}^{-1}$.}
\end{figure}

Based on the displacement field \eqref{e:steady_state_displ}, we obtain the total relative loss of volume in response to the osmotic shock,
\begin{equation}\label{e:volume_loss}
\frac{\Delta V}{V}=\frac{1}{|\Omega|}\int_{\Omega}\text{tr}(\mathbb{E}_{\text{eq}}) d\underline{X}=\frac{3}{R_m^3}\int_{0}^{R_m}R^2\text{tr}(\mathbb{E}_{\text{eq}}) dR=-\frac{\Pi_e}{K_{\text{active}}(\alpha,\beta,K_d,G,P_a)}
\end{equation}
where, due to its length, we omit here the explicit expression of $K_{\text{active}}$  computed using formula \eqref{e:steady_state_displ}. With these estimated parameters, we obtain $K_{\text{active}}\simeq 32$ kPa which is close to the value $K_{\text{exp}}\simeq 28\in (21,34)$ kPa measured experimentally (See Fig.~\ref{fig:compression_diffusion}~(a)) according to the protocol described in \ref{sec:MCS_compression_exp}. The parenthesis denote the 95$\%$ confidence interval of the fit.

\section{Water relaxation}\label{sec:water_relaxation}

The active poroelastic theory \eqref{e:problem_poro_elast_simpl} also enables us to predict the dynamics of the water percolation in the MCS pores after the application of an osmotic pressure. Indeed, we have
\begin{equation}\label{e:water_relax_Fa}
\frac{\partial\text{tr}(\mathbb{E})}{\partial t}-\frac{\kappa}{\mu R^2}\frac{\partial}{\partial R}\left( R^2\frac{\partial\delta p}{\partial R}\right) =0 \text{ with }\frac{\partial\text{tr}(\mathbb{E})}{\partial t}= 3 \int_R^{R_m} v^3\left[R f'(R)+3 (f(R)-f(v))\right]\frac{\partial^2 \delta p(v,t)}{\partial v\partial t} \, dv,
\end{equation}
which is a non-local equation on $\delta p$. In the absence of active stress,
$$f(R)=\frac{1}{3(4G+3K_d)}\left( \frac{1}{R^3}-\frac{1}{R_m^3}\right) $$
and \eqref{e:water_relax_Fa} reduces to \eqref{e:problem_poro_elast_lin}. When activity is present, \eqref{e:water_relax_Fa} is not a simple diffusion of the interstitial water percolating out of the MCS. Denoting by $p'(R,t)=\partial_R\delta p$ the spatial derivative of pressure we derive from \eqref{e:water_relax_Fa}:
\begin{equation}\label{e:water_relax_2}
\frac{3 R^3 \partial_t p'(R,t) (4 G+3 K_d+3 P^a_r(R))+9 \partial_RP^a_r(R) \left(\int_R^{R_m} v^3 \partial_tp'(v,t) \, dv\right)}{R^3 (4 G+3 K_d+3 P^a_r(R))^2}=\frac{\kappa}{\mu}\left(\partial_{RR} p'(R,t)+\frac{2 \partial_{R} p'(R,t)}{R}-\frac{2  p'(R,t)}{R^2} \right). 
\end{equation}
Introducing the re-scaled spatial variable $x=R/R_m$ in the above equation, it is clear that its relaxation time $\tau$ scales as
$$\tau=\frac{R_m^2\mu}{\kappa}q\left( K_d+\frac{4}{3}G,P_a\right),$$
where $q$ is an unknown function that has the dimension of the inverse of a stress and reduces to 
$$q\left( K_d+\frac{4}{3}G,0\right)=\frac{1}{K_d+4G/3}$$
in the absence of active stress. We show on Fig.~\ref{fig:compression_diffusion}~(b) that the predicted scaling with the square of the MCS radius matches the experimental data. We outline in \ref{sec:timescale_pressure} a classical method to obtain a rigorous numerical expression of $q$. Instead, here, to roughly estimate the influence of the active stress on $q$,  we  inject a diffusive ansatz into \eqref{e:water_relax_2}:
$$p_{\text{eff}}'(R,t)=-\frac{R \text{e}^{-\frac{R^2}{4 D_a t}}}{2 D_a t^{5/2}}$$
where $D_a$ is an effective diffusion coefficient. For the initial response  when $D_a t\ll R_m^2$ that corresponds to most of the spheroid volume loss, such an ansatz leads to the relation
$$ \frac{4 G}{3}  +   K_d+  P_a\left( 1- \left(\frac{R_m}{R}\right)^{\beta }\right) =\frac{D_a \mu}{\kappa},$$ 
which obviously cannot be fulfilled pointwise (since the ansatz does not exactly solve \eqref{e:water_relax_2}) but can be considered in average in the whole MCS leading to:
$$D_a=\frac{\kappa(4G/3+K_d+\beta P_a/(\beta-3))}{\mu}.$$
As,
$$\frac{1}{R_m^3}\int_0^{R_m}R^2P_r^a(R)dR=\frac{\beta P_a}{\beta-3}\geq 0,$$
this expression corresponds to considering an homogeneous active stress at the level of $f$ in \eqref{e:water_relax_2}. In this effective formula, the  active stress in the MCS therefore increases the interstitial water hydrodynamic diffusivity by adding up to an active contribution to $D_{\text{passive}}$. However, as it relies on a rough ansatz, such a formula is not quantitative and we only retain the fact that the order of magnitude of $q$ is $1/K_d$. 

Based on this estimate, with the measurement of $D_{\text{exp}}$, we can roughly approximate the MCS permeability as $\kappa\simeq \mu D_{\text{exp}}/K_d\simeq 1 \times 10^{-18}\text{ m}^2$ which is comparable to some (broadly distributed) previous measurements \citep{swartz2007interstitial} but differs from others performed with less cohesive aggregates produced with a different cell type \citep{Tran2018}. 

\section{ECM rheology and volume exclusion of cells can explain the MCS mechanical properties }\label{sec:composite}

In this section, we motivate the fact that the effective rheological coefficients of the MCS, $K_d$ and $\kappa$ can be interpreted as stemming from the ECM properties while cells are  simply impermeable and incompressible objects which are only responsible for volume exclusion. \tr{In general, even at a short timescale of few minutes, individual cells do respond to a hyperosmotic shock by changing their volume through the efflux of water from their cytosol \citep{cadart2019physics}. However we do not expect such behavior in the present context for two reasons. First, we use osmolytes that are too large to permeate the MCS pores such that they could only affect cells at the periphery of the MCS. Second and more important, the magnitude of the applied osmotic shocks in the present experiments are typically less than $5$ kPa representing a concentration of osmolites of few milimolars, negligible compared to the typical ionic concentrations of culture media of few hundreds milimolars. Such shocks are thus by far too small to have any influence on individual cells as experimentally demonstrated by \cite{monnier2016effect}. Still the volume of cells could well vary indirectly due to the mechanical stress in the ECM provoked by the shock, motivating our analysis to investigate whether our assumption of cells as impermeable and incompressible in the present context is in agreement with the experimental data.}

 As interstitial ECM is difficult to characterize in-situ, we use  matrigel (MG) beads to roughly estimate the rheological properties of ECM. MG is an ECM proxy secreted by EHS mouse sarcoma \citep{kleinman1982isolation}. Consistently with native ECM, large Dextran molecules were also excluded from microbeads made of MG  suggesting an equivalent effective permeability (See~\ref{sec:poro_MG} and \ref{sec:poro_ECM}). Beads of MG are typically passive poroelastic materials to which we can apply the results of Section~\ref{sec:passive}.

\begin{figure}[h!]
\centering
\includegraphics[width=1.0\textwidth]{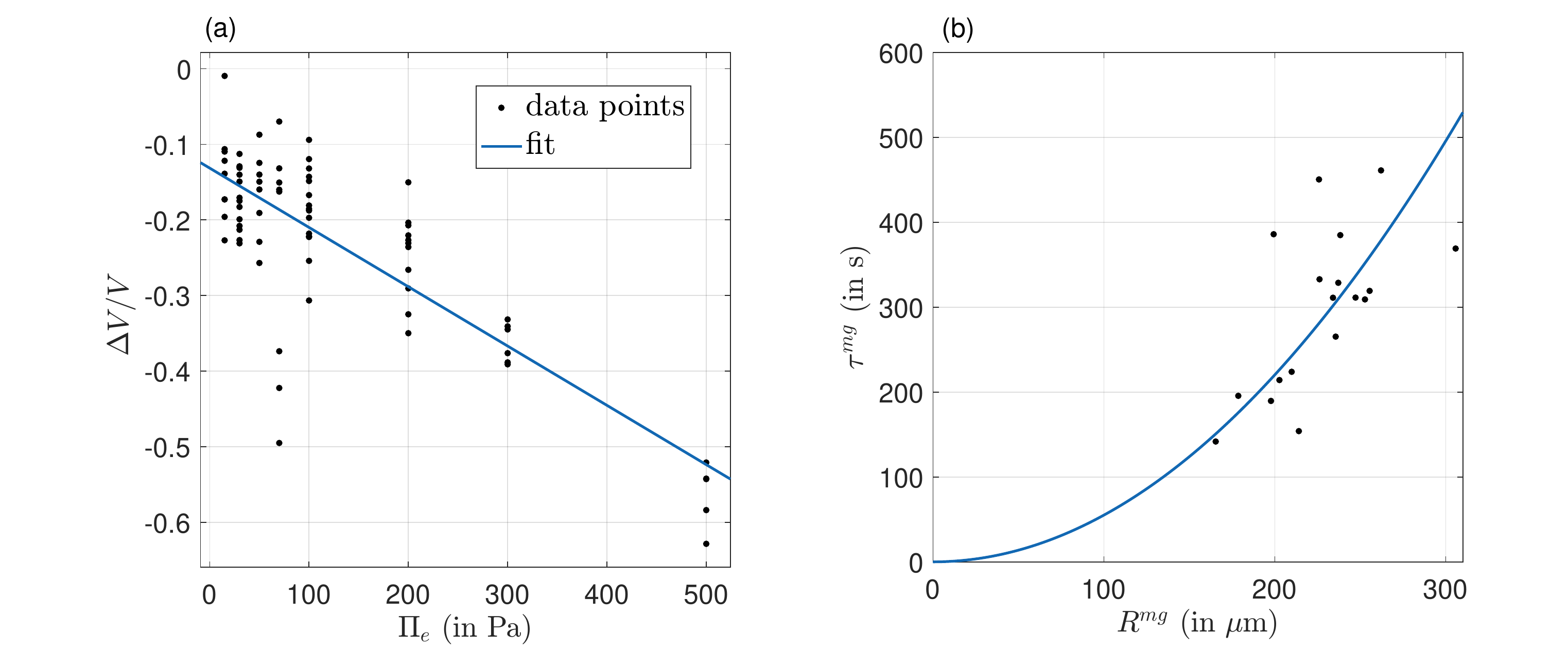}
\caption{\label{fig:MGcompression_diffusion} \textbf{ Poroelastic properties of MG beads.} (a) Compression of MG beads as a function of osmotic pressure. The linear fit is $\Delta V/V=-\Pi_e/K_d^{\text{mg}}+\text{offset}$ where $K_d^{\text{mg}}\simeq 1270 \in (1060,1490)$ Pa and $\text{offset} \simeq-0.13\in(-0.1,-0.15)$. (b) Relaxation time of MG beads as a function of their initial radius. The quadratic fit corresponds to $\tau^{\text{mg}}=(R^{\text{mg}})^2/(6D^{\text{mg}})$ where $D^{\text{mg}}\simeq 3\in(2.6,3.4)\times 10^{-11}\text{m}^2\text{s}^{-1}$. }
\end{figure}

\begin{figure}[h!]
\centering
\includegraphics[width=.7\textwidth]{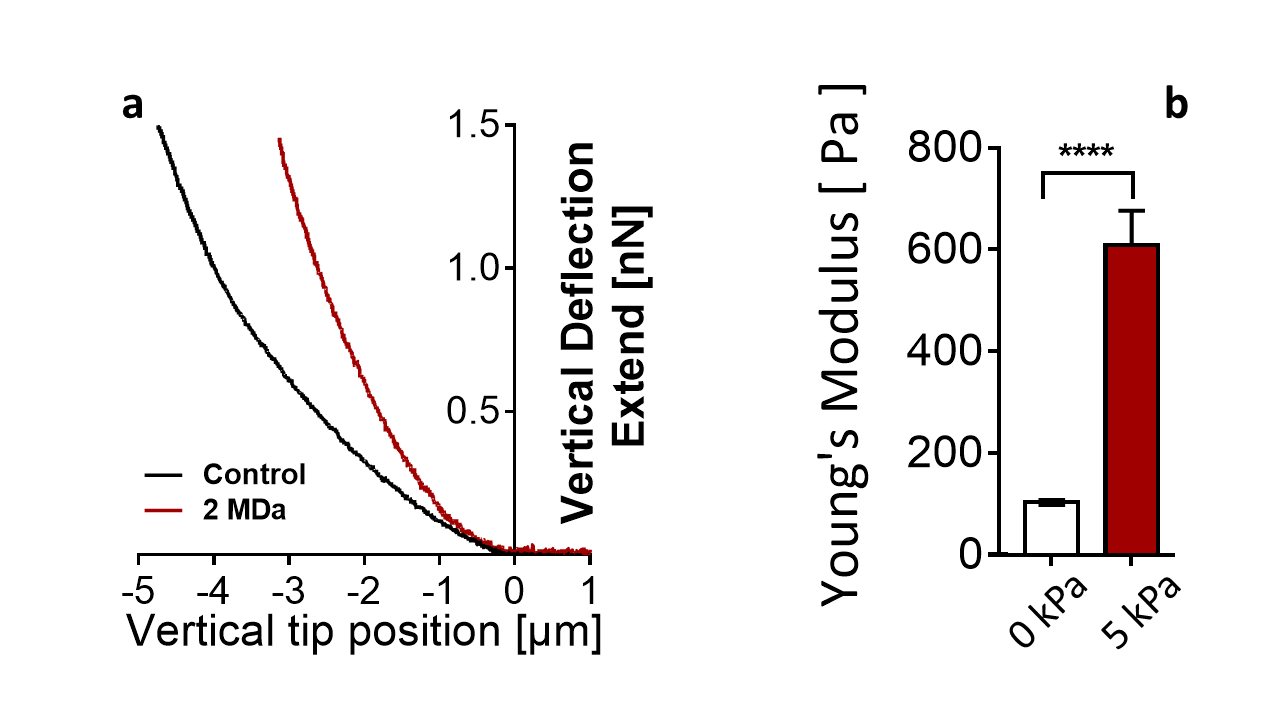}
\caption{\label{fig:MG_Shear} \tr{\textbf{Matrigel Young modulus measured by Atomic Force Microscopy.}
(a) Force/Indentation response of a MG layer before (black curve) and after an osmotic compression of 5 kPa (red curve). The curve displays the indentation force F as a function of the vertical tip indentation.
(b) Qualification of the stiffening behavior of MG under osmotic compression: the Young modulus increases by 6-fold under $\Pi_e=5$kPa.  In (b), error bars represent the standard deviation. A T-test was used for statistical significance. Experiments were repeated on three different gels per condition. }}
\end{figure}

\paragraph{Drained modulus}
We thus begin by estimating the bulk modulus $K_d^{\text{mg}}$ of MG beads by applying osmotic pressure shocks (experimental protocol described in \ref{sec:MG_beads_prepa}). See Fig.~\ref{fig:MGcompression_diffusion}~(a).
By fitting the data with a linear curve, we obtain  $K_d^{\text{mg}}\simeq 1270 \in (1060,1490)$ Pa. Note the presence of an offset in the compression data probably due to ionic affinities of some MG components  when the Dextran solution used to perform the osmotic shock is introduced.
To complement this measurement, we also perform AFM compression of MG layers to estimate their shear modulus \tr{(See the experimental protocol in \ref{Sec:Young_Shear}). We find that the Young modulus of MG layer is about 100 Pa and stiffens under osmotic compression to reach values of about 600 Pa for $\Pi_e=5$ kPa. See Fig.~\ref{fig:MG_Shear}.}
\tr{This leads to a shear modulus $G^{\text{mg}}\simeq 100$ Pa  which is small compared to the bulk modulus.}

If we now suppose that cells within the MCS  are incompressible \tr{-their osmotic bulk modulus is $\sim 800$ kPa $\gg K_d^{\text{mg}}$ as measured by \cite{monnier2016effect}-} inclusions coated by MG, we can  deduce a bulk modulus of the MCS from the Hashin-Shtrikman upper bound \citep{hashin1963variational, hashin1962elastic}, which we expect to be a good approximation as the contrast between the bulk modulus on cells and MG is large \citep{milton2002theory}
$$K_d\simeq K_d^{\text{mg}}+\frac{1-n_0}{n_0}\left( K_d^{\text{mg}}+\frac{4G^{\text{mg}}}{3}\right),$$
where again $n_0$ is the average MCS porosity prior to osmotic compression.

\begin{figure}[h!]
\centering
\includegraphics[width=0.7\textwidth]{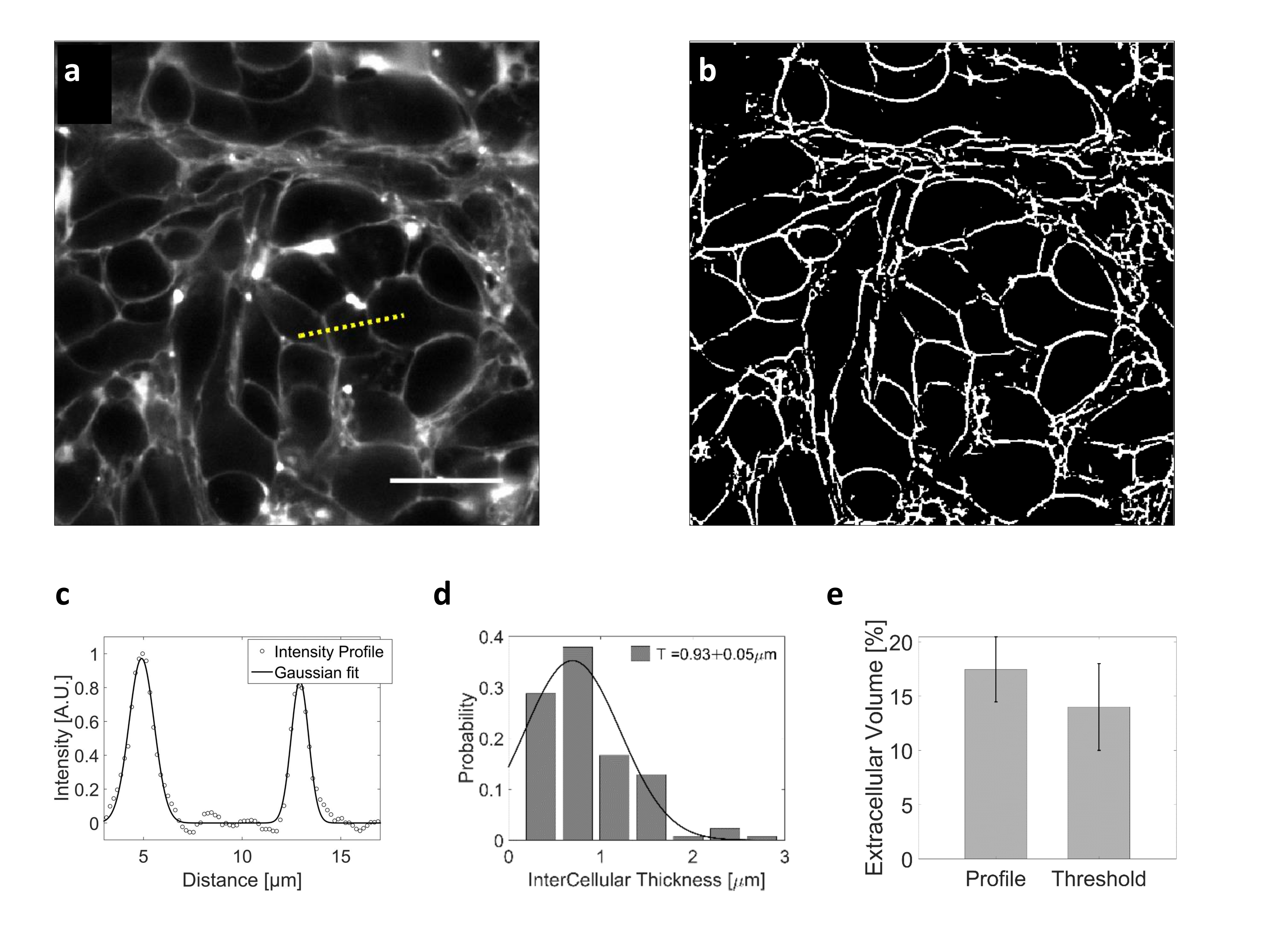}
\caption{\label{fig:ECM_Ratio} \tr{\textbf{Extra-cellular volume fraction estimation.} (a) Thickness method: confocal section of an uncompressed MCS, the extracellular space of which is filled with sulforhodamine-B, a hydrophilic fluorophore that does not enter the cells. (c) The intensity profile across two extracellular layers. The width of intercellular space is computed by fitting the intensity to a gaussian profile and shown on panel (d). (b) Threshold methods:  A threshold is applied to image (a)  to evaluate the ratio between extracellular (white) and intracellular (black) space. Depending on the choice of the threshold, the ratio of extracellular space varies by $\pm5\%$. (e) Summary of the results obtained using the two methods. Whereas the experimental error is large, the observations show that the extracellular space represents about 15$\%$ of the total MCS volume.}}
\end{figure}

\tr{To evaluate $n_0$, we supplement the culture medium with sulforhodamine-B, a hydrophilic fluorophore that stains the extra-cellular space without penetrating the cells. From confocal sections of MCS (Fig.~\ref{fig:ECM_Ratio}~a) we determine the thickness of the thin layer between two adjacent cells. By fitting the intensity profile to a Gaussian distribution (Fig.~\ref{fig:ECM_Ratio}~c), and taking into account that the instrumental function (resolution 270 nm) broadens the profile, we estimate the extracellular layer to 0.9$\pm$0.1 $\mu$m  (histogram in Fig.~\ref{fig:ECM_Ratio}~d; N=132). With an average cell diameter of 20 $\mu$m, we evaluate that the fraction of extracellular space (i.e. porosity) is approximately $n_0 = V_m/V_0 = 14\pm5 \%$.
An alternative method to evaluate the fraction $n_0$ is to threshold the fluorescence intensity measured in a confocal section (Fig.~\ref{fig:ECM_Ratio}~a). From the number of white pixels in the images after thresholding (Fig.~\ref{fig:ECM_Ratio}~b), we estimate  that the fraction of extracellular space is $ n_0 = 14\pm4 \%$. The experimental uncertainty is due to the empiric choice of parameters used in the threshold process. Fig.~\ref{fig:ECM_Ratio}~e compares the results obtained with the two methods.}

We then obtain an estimate of $K_d\simeq 10$~kPa which is of the same order of magnitude as the one suggested by our theory and experimental measurements presented above ($K_d\simeq 30$ kPa). A quantitative match is not necessarily expected as MG and ECM in situ in the spheroid can have bulk moduli that differ by a factor of three. 

\paragraph{Permeability} As in the MCS case, we can also measure the characteristic time associated to water percolation needed to compress MG beads to estimate the diffusion coefficient $D^{\text{mg}}\simeq 3\in(2.6,3.4)\times 10^{-11}\text{m}^2\text{s}^{-1}$. See Fig.~\ref{fig:MGcompression_diffusion}~(b). Since $\kappa^{\text{mg}}=\mu D^{\text{mg}}/(K_d^{\text{mg}}+4/3G^{\text{mg}})$, we estimate the MG permeability to be $\kappa^{\text{mg}}=2.3\times 10^{-17} \text{m}^2$. This value is of the correct order of magnitude as permeability scales with the square of the characteristic mesh size which is estimated to be roughly $10$ nm both for MG beads and for ECM in the MCS (See \ref{sec:poro_MG} and \ref{sec:poro_ECM}).

Considering the cells as almost impermeable compared to the extra-cellular space, we can estimate the MCS permeability according to the classical Maxwell result as,
$$\kappa=\frac{2n_0}{3-n_0}\kappa^{\text{mg}}\simeq 2 \times 10^{-18}\text{m}^2,$$
which is again of the same order of magnitude as our estimation based on measurements on MCS ($\kappa\simeq 1 \times 10^{-18}\text{ m}^2$).

\section{Conclusion}\label{sec:conclusion}
In this paper, we have presented an effective  active poroelastic theory to model the response of a young (devoid of a necrotic core and vasculature) MCS subjected to an osmotic compression on the timescale of minutes to hours. At these timescales, the material response is not active \emph{per se} but the poroelastic deformation is modified by the existing active stress and interstitial pore pressure stemming from cellular turnover in the MCS. 

Our theory qualitatively captures the solid strain and stress distributions within the aggregate upon an osmotic shock. In particular, we rationalize the fact that while the displacement induced by the shock vanishes at the center of the MCS, the associated strain is large. \tr{As we explain in Section~\ref{sec:stress_strain}, the theory also assumes a form of the total active stress prior to the osmotic compression that is consistent with cutting experiments \citep{stylianopoulos2012causes, colin2018experimental, guillaume2019characterization}. The active stress  is considered to be fixed during the osmotic shock response.} Comparing such a theory with experimental results, we have estimated two key passive phenomenological coefficients that control the compression: the drained bulk modulus of the MCS and the hydraulic diffusion of water permeation in the spheroid pores and explained how they are redressed by activity.  Additionally, we suggest that both the effective drained modulus and the MCS permeability can be attributed to ECM properties corrected by volume exclusion of incompressible and impermeable cells.

In this framework, the ECM acts as a sensor through which the global MCS osmotic compression is transformed into a permanent mechanical solid stress that acts as a biochemical signal impacting cell proliferation. While our preliminary results \citep{Dolega488635} indicate that the MCS rheology and, in particular, the state of compression of the ECM affect cellular fate within the MCS on a long time scale, cellular turnover will in turn modify the active stress leading to an emergent hydrodynamic diffusion and mechanical stress within the MCS, that can be fundamental features controlling the cells collective behaviors  \citep{recho2019theory}. We therefore anticipate that mechanical theories aiming at capturing such a state should further account for the presence of the ECM and its coupling with cell proliferation.

\section{Acknowledgments}
We  thank J. Prost and F. J\"{u}licher for drawing our attention to the potential impact of the poroelasticity on the rheology of multicellular aggregates.
This work was supported by the Agence Nationale pour la Recherche (Grant ANR-13-BSV5-0008-01), by the Institut National de la Sant\'e et de la Recherche M\'edicale (Grant ``Physique et Cancer'' PC201407) and by the Centre National de la Recherche Scientifique (grant Momentum and grant MechanoBio 2018).  This work has been partially supported by the LabeX Tec 21 (Investissements d'Avenir: grant agreement No. ANR-11-LABX-0030).

\appendix 
\renewcommand\thefigure{\arabic{figure}}

\section{\tr{Thermodynamic foundations of the model}}\label{sec:thermo_found}

\tr{In its actual configuration, we model the MCS as a continuum object composed of two  phases in a representative volume element: the network of cells with an actual volume fraction $n_s(\underline{x},t)$ and the extra-cellular fluid with an actual volume fraction $n(\underline{x},t)$. The saturation constraint imposes that $n_s+n=1$. We can directly incorporate this condition in the mass balance equations:
\begin{equation}\label{e:mass_bal_solid}
\begin{array}{c}
\partial_{t\vert x}(\rho n)+\nabla_x.(\rho n\underline{v}_e)=s\\
\partial_{t\vert x}(\rho_s(1-n))+\nabla_x.(\rho_s (1-n)\underline{v}_s)=-s\\
\end{array}.
\end{equation}
In \eqref{e:mass_bal_solid}, the time and space derivatives are the Eulerian ones -related to the actual configuration- this is why we do not use the same notations as in the main paper where the derivatives are related to the initial Lagragian frame. The source term $s$ in this Eulerian frame is related to the original Lagragian source term $S$ by the relation $S=Js$. The fluid velocity in the laboratory frame is denoted $\underline{v}_e$ and the velocity of the cell network is $\underline{v}_s=\partial_t\underline{u}$.} 

\tr{The force balance in each phases takes the form
$$-\nabla_x(n p)=\underline{f}_e \text{ and }
\nabla_x.((1-n)\mathbb{\Sigma}_s)=\underline{f}_s,$$
where the Cauchy stress in the cell network phase is denoted $\mathbb{\Sigma}_s$ and the interaction forces between the fluid and network phases satisfy $\underline{f}_e+\underline{f}_s=0$ as no external force is acting on the MCS. Again this constraint is directly accounted for in the global stress balance:
\begin{equation}\label{e:force_bal_2}
\nabla_x.\mathbb{\Sigma}_t=0,
\end{equation}
where the total Cauchy stress reads $\mathbb{\Sigma}_t=(1-n)\mathbb{\Sigma}_s-n p\mathbb{I}$. Note the absence of any Lagrange multiplier in \eqref{e:force_bal_2} as no incompressibility condition is a priori imposed. With the classical definition of the Piola-Kirchhoff stress $\mathbb{P}=J\mathbb{\Sigma}_t\mathbb{F}^{-T}$, \eqref{e:force_bal_2} is strictly equivalent to \eqref{e:force_bal}.}

\tr{Next, to investigate the constitutive behaviour of the spheroid, we express its internal energy dissipation as 
$$\Theta=\frac{\delta W}{\delta t}-\frac{\delta G}{\delta t},$$ 
where we have assumed a constant temperature and $\delta W/\delta t$ is the power exerted by external forces on the system and $\delta G/\delta t$ the variation of the system free energy. The second principle of thermodynamics imposes that $\Theta\geq 0$. See \citep{Recho2014} for more details.}

\tr{The external power can be computed as 
$$\frac{\delta W}{\delta t}=\int_{\partial\omega_t}\left( (1-n)\mathbb{\Sigma}_s\underline{n}.\underline{v}_s-np\underline{n}.\underline{v}_e\right) d\underline{x}=\int_{\omega_t}\nabla_x.((1-n)\mathbb{\Sigma}_s\underline{v}_s-np \underline{v}_e)d\underline{x},$$
where $\underline{n}$ is the outward unit vector normal to the MCS surface. Introducing the relative velocity of the fluid with respect to the solid 
$$\bar{\underline{v}}_{e}=\underline{v}_{e}-\underline{v}_s$$
and using the momentum balance law \eqref{e:force_bal_2}, we finally obtain,
$$\frac{\delta W}{\delta t}=\int_{\omega_t}\left( \mathbb{\Sigma}_t:\nabla \underline{v}_s-\nabla_x.(n p\bar{\underline{v}}_{e})\right) d\underline{x},$$
where $:$ denotes the Hadamard product. As the balance of torques implies that $\mathbb{\Sigma}_t$ is symmetric,  $\mathbb{\Sigma}_t:\nabla_x \underline{v}_s=\mathbb{\Sigma}_t:\mathbb{D}$, where $\mathbb{D}=(\nabla_x \underline{v}_s+(\nabla_x \underline{v}_s)^T)/2$ is the symmetric part of the velocity gradient tensor and we obtain,
\begin{equation}\label{e:work_rate}
\frac{\delta W}{\delta t}=\int_{\omega_t}\left( \mathbb{\Sigma}_t:\mathbb{D}-\nabla_x.(n p\bar{\underline{v}}_{e})\right) d\underline{x}.
\end{equation}}

\tr{We assume that the free energy depends on three variables 
$$G=\int_{\Omega} g(\mathbb{E},m,\Lambda)d\underline{X},$$
where the two classical mechanical variables are $\mathbb{E}$, the Cauchy-Green strain tensor and $m=\rho n J$ the mass of external fluid per unit reference volume and $\Lambda=\rho_s (1-n) J\zeta$ is a biochemical variable that fuels the cell active processes (hence proportional to the cell mass). Therefore $\zeta$ is the same variable per unit mass and can be associated with the extend of ATP hydrolysis as it is classically done in the theory of active gels \citep{kruse2005generic}. We can then derive:
$$\frac{\delta G}{\delta t}=\int_{\omega} J^{-1}\left( \partial_{\mathbb{E}}g:\partial_t\mathbb{E}+\partial_{m}g\partial_tm+\partial_{\Lambda}g\partial_t\Lambda\right)d\underline{x} $$
As we have from \eqref{e:mass_bal_solid} that,
$$J^{-1}\partial_tm=s-\nabla_x.(\rho n \bar{\underline{v}}_{e} )\text{ and } J^{-1}\partial_t\Lambda=-s\zeta+\rho_s(1-n)\partial_t\zeta,$$
we obtain the final expression for the dissipation
$$\Theta=\int_{\omega_t}\left( \left( \mathbb{F}^{-1}\mathbb{\Sigma}_t\mathbb{F}^{-T}-\frac{\partial_{\mathbb{E}}g}{J}\right) :\partial_t\mathbb{E}+\nabla_x.\left( (\rho\partial_{m}g-p)\bar{\underline{v}}_{e} \right)-n\bar{\underline{v}}_{e}\nabla_x(\rho\partial_{m}g)+s\left(\zeta\partial_{\Lambda}g-\partial_{m}g \right) -\rho_s(1-n)\partial_{\Lambda}g\partial_t\zeta\right) d\underline{x}.$$}

\tr{Following the classical poroelastic theory \citep{coussy2004poromechanics}, we then assume that there is no dissipation in the solid skeleton and in the bulk of the permeating fluid leading to the relations:
$$\mathbb{\Sigma}_t=\frac{1}{J}\mathbb{F}\partial_{\mathbb{E}}g\mathbb{F}^T\text{ and }p=\rho\partial_mg,$$
equivalent to \eqref{e:thermo_rel}. The dissipation then reduces to 
\begin{equation}\label{e:dissip_final}
\Theta=\int_{\omega_t}\left( -n\bar{\underline{v}}_{e}\nabla_xp+s\left(\zeta\partial_{\Lambda}g-\frac{p}{\rho} \right) -\rho_s(1-n)\partial_{\Lambda}g\partial_t\zeta\right) d\underline{x}.
\end{equation}
Following the close-to-equilibrium Onsager framework, the positivity of the dissipation related to the percolation of fluid in the interstitial space is insured by the linear relation (i.e. Darcy's law)
$$n\bar{\underline{v}}_{e}=-\mathbb{k}\nabla_x p,$$    
where $\mathbb{k}$ is the positive definite permeability matrix in the actual configuration. Notice that as for any vector field, $\nabla()=\mathbb{F}^T\nabla_x()$ and $\nabla.()=J\nabla_x.(J^{-1}\mathbb{F}())$,  this relation leads to the conservation law
$$\partial_tm-\nabla.(\rho J\mathbb{F}^{-1}\mathbb{k}\mathbb{F}^{-T}\nabla p )=Js=S.$$
As the permeability in the reference configuration, prior to any deformation, is given by $\mathbb{k}=\frac{1}{J}\mathbb{F}\mathbb{k}_0\mathbb{F}^T$, the above equation reduces to \eqref{e:const_be} where we  assume that $\mathbb{k}_0=\kappa \mathbb{I} $ is isotropic and $\kappa$ is a fixed quantity that does not depend on $m$.}

\tr{The last two terms in \eqref{e:dissip_final} represent the dissipation related to the interconnected processes of cell division/death and energetic fueling of the active system. As we do not make explicit the dependence of $g$ on $\Lambda$ in our phenomenological model, these processes are not thermodynamically specified.  It therefore remains to choose a certain form of $g$ to close the mechanical problem under investigation. In the initial configuration the mass of interstitial fluid per reference volume element is $m(\underline{X},0)=m_0(\underline{X})$ and by definition $\mathbb{E}=0$. Assuming small deviations from these quantities upon the osmotic perturbation, we approximate $g$ by its expansion  up to quadratic order in these two variables. If the spheroid is isotropic, the symmetry of the problem implies that
$$g=g(0,m_0,\Lambda)+\frac{\partial g}{\partial \mathbb{E}}(\Lambda)\vert_{\mathbb{E}=0,m=m_0}:\mathbb{E}+\frac{\partial g}{\partial m}(\Lambda)\vert_{\mathbb{E}=0,m=m_0}\delta m+\frac{K_u(\Lambda)}{2}\text{tr}(\mathbb{E})^2+G(\Lambda) \mathbb{E}:\mathbb{E}-\zeta(\Lambda) \text{tr}(\mathbb{E})\delta m+\frac{\chi(\Lambda)}{2}\delta m^2,$$
in which $\delta m=m-m_0$ and the active stress and interstitial pressure prior to the osmotic shock are:
$$\mathbb{P}_a=\frac{\partial g}{\partial \mathbb{E}}(\Lambda)\vert_{\mathbb{E}=0,m=m_0}\text{ and }p_a=\rho\frac{\partial g}{\partial m}(\Lambda)\vert_{\mathbb{E}=0,m=m_0}.$$}

\tr{At the timescale of the response of the MCS to the osmotic compression (few tens of minutes), we consider that $\Lambda(\underline{X})$ is fixed and does not depend on time. This leads to an active pre-stress $\mathbb{P}_a(\underline{X})$ and interstial pressure $p_a(\underline{X})$ that are also time independent. In general, the rheological coefficients that characterize the   second order of the free energy $K_u$, $G$, $\zeta$ and $\chi$ should also depend on space through $\Lambda$ but for simplicity, we assume here that they are constants.}

\section{Spatial distribution of the initial pore pressure}\label{sec:pa}

In this Section, we reconstruct the initial hydrostatic pressure distribution by considering \eqref{e:conditions_pa} with the special form of the (interstitial fluid) source term:
$$S(R)=S_a \left(1-\left(\frac{\alpha }{2}+1\right) \left(\frac{R}{R_m}\right)^{\alpha }\right)$$
with $\alpha\geq 0$ and $S_a>0$. As a result, $S(R_m)<0$ is negative a the MCS surface since interstitial fluid is uptaken by growing cells at this location and $S(0)>0$ since cells die in the core. The steady slow (associated to the cell proliferation timescale) flow of water with respect to the cellular flow is $v_w=-(\kappa/\mu)\partial_Rp_a$. 
\begin{figure}[h!]
\centering
\includegraphics[width=0.8\textwidth]{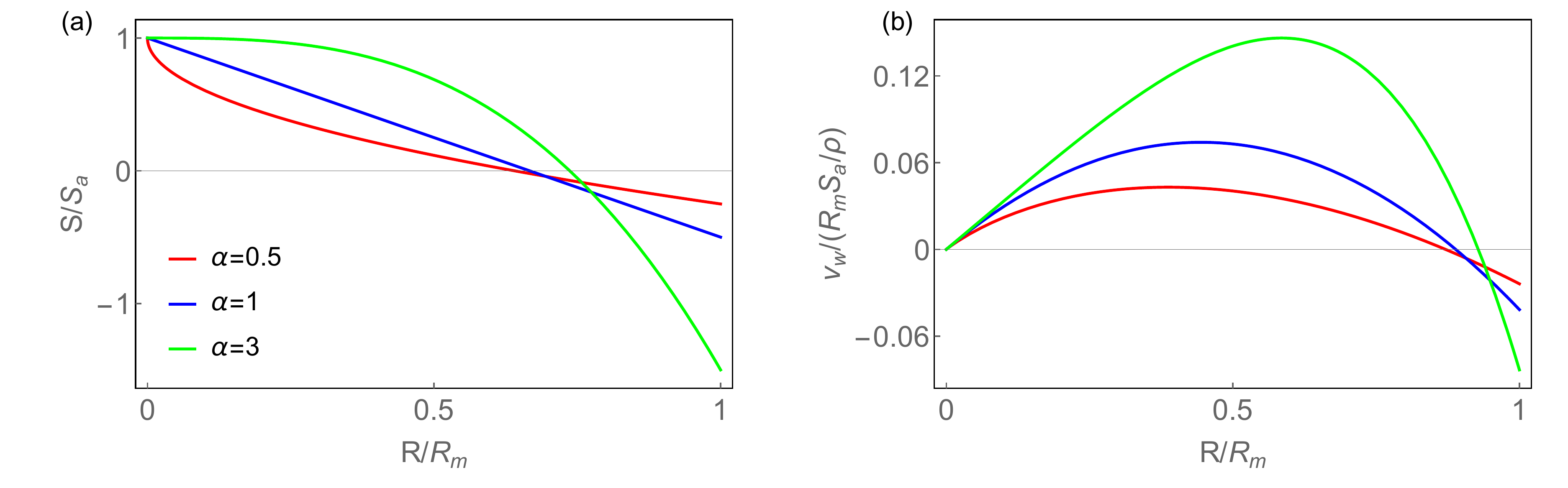}
\caption{\label{fig:pa} \textbf{Active interstitial flow in the MCS.} (a) Spatial distribution of the source term for various choice of $\alpha$. (b) Associated velocity of the interstitial fluid flow relative to the MCS solid phase. }
\end{figure}
\FloatBarrier

Imposing the symmetry condition $v_w(0)=0$, the solution of \eqref{e:conditions_pa} is given by 
$$p_a(R)=\delta \Pi_a+\frac{\mu  S_a \left(R^2 \left(-\alpha +3 \left(\frac{R}{R_m}\right)^{\alpha }-3\right)+\alpha  R_m^2\right)}{6 (\alpha +3) \kappa  \rho }$$
and leads to,
$$v_w(R)=\frac{R S_a }{6\rho} \left(2-\frac{3 (\alpha +2) \left(\frac{R}{R_m}\right)^{\alpha }}{\alpha +3}\right),$$
which is displayed on Fig.~\ref{fig:pa}. Such a velocity field is opposite to the one measured for cells inside the MCS \citep{delarue2013mechanical}.

\section{Typical timescale of the pressure relaxation in the MCS.}\label{sec:timescale_pressure}

To quantitatively understand the variations of $q$ introduced in Sec.~\ref{sec:water_relaxation}, a classical method is to project $p'(R,t)$ on an Hilbert basis satisfying the boundary conditions ($\delta p(R_m,t)=0$ and $p'(R_m,t)=0$); typically Bessel functions
$$\forall k\geq 0,\, p'_k(R)=\frac{\sqrt{2} \left(\pi  k R \cos \left(\frac{\pi  k R}{R_m}\right)-R_m \sin \left(\frac{\pi  k R}{R_m}\right)\right)}{\pi  k R^2 \sqrt{R_m}}$$
are  good candidates as they form an orthonormal basis for the $L^2$ scalar product and diagonalize the left handside of \eqref{e:water_relax_2}. Then we express
 $$p'(R,t)=\sum_{k=0}^{\infty}\text{e}^{-\lambda_kt}p'_k(R).$$
The same basis is used to project $P^a_r$ to transform \eqref{e:water_relax_2} into an infinite linear system solving for $\lambda_k$. The solution of such a problem can be approximated numerically and a relaxation function is then given by:
$$||p'||_2^2(t)=\int_0^{R_m}R^2p'(R,t)^2dR=\sum_{k=0}^{\infty}\text{e}^{-2\lambda_k t}.$$
The average of that function can be related to the relaxation time $\tau$ introduced in Sec.~\ref{sec:water_relaxation},
$$\tau=\underset{T\rightarrow \infty}{\lim}\frac{1}{T}\int_0^T||p'||_2^2(t)dt=\sum_{k=0}^{\infty}\frac{1}{2\lambda_k}.$$
The scaling of $\tau$ as
$$\tau=\frac{R_m^2\mu}{\kappa}q\left( K_d+\frac{4}{3}G,P_a\right),$$
directly follows from the spatial non-dimensionalization of \eqref{e:water_relax_2} which implies that $\kappa/(R_m^2\mu)$ is a common factor of all the $\lambda_k$.

\section{Cell culture, MCS formation and osmotic compression.}\label{sec:MCS_compression_exp}
CT26 (mouse colon adenocarcinoma cells, ATCC CRL-2638; American Type Culture Collection) were cultured under 37$^{o}$C, 5$\%$ $\text{CO}_2$ in DMEM supplemented with 10$\%$ calf serum and 1$\%$ antibiotic/antimycotic (culture medium). 
MCS were prepared on an agarose cushion in 96 well plates at the concentration of 500 cells/well and centrifuged initially for 5 minutes at 800rpm to accelerate aggregation. After 2 days, Dextran with a Molecular Weight of 2 MDa (Sigma-Aldrich, D5376-100G) was added to the culture medium to exert osmotic pressure, as previously described in \cite{monnier2016effect}, at a concentration of 55 g/L to exert $\Pi_e =5 $kPa. 
Images were analysed using the Image J plugin developped in \cite{ivanov2014multiplexing}. 

\section{\tr{Protocols of the AFM measurments of Young moduli of MG and MCS}} \label{Sec:Young_Shear}

\subsection{MG}
To measure the evolution of MG Young modulus with the compressive stress, we prepared a MG film, the thickness of which was about $500\mu$m, on a plastic petridish. The film was maintained flat using a hydrophobic plastic film and was kept at $37^{\circ}$C for 20 minutes to favor MG polymerization. Eventually, the hydrophobic film was carefully removed. MG gels were prepared at least 1.5 hours before the experiment and remained in PBS to avoid gel swelling during AFM measurements (JPK Nanowizard II mounted on inverted microscope C. Zeiss, Observer D1).  We used the AFM cantilever  (Bruker, MLCT pyramid shape, $k=0.013$ N/m) to measure Young modulus. The approach speed was set to 1 $\mu$m/s.

\subsection{MCS}\label{Appendix:shear_MCS}
To determine the shear modulus of the MCS, we used  the same AFM as for MG with a stiffer cantilever (Nanosensors, TL-NCL-10, tipless, Length $L=250\mu$m, $k = 30$ N/m). The MCS radii were in the range 100-430 $\mu$m. The indentation speed varied between 20 and 40 $\mu$m/s and the MCS relaxed with a typical time $\tau = 59\pm 3s$ (standard error of the mean, N=4). This value corresponds to the poroelastic timescale, due to water percolation.

\section{\tr{protocol of the} MG beads preparation and compression}\label{sec:MG_beads_prepa} 
MG beads were prepared using the vortex method. Oil phase of HFE-7500/PFPE-PEG (1.5 $\%$w/v) was cooled down to 4$^o$C. For 400 $\mu$L of oil, 100$\mu$L of MG was added. The solution was vortexed for 20 seconds and subsequently kept at 37$^\circ$C for 20 minutes for polymerization. Beads were transferred to the PBS phase by washing out the surfactant phase with pure HFE-7500 oil. To compress polimerized Matrigel beads, PBS was enriched with 2MDa Dextran at the concentration of 56 g/l, which corresponds to an osmotic pressure of $\Pi_e=5$ kPa. Images were taken just before, and  dextran was added 45 minutes after. Volume decrease was measured for 10 different beads.

\section{\tr{Pore size} of MG beads}\label{sec:poro_MG}
\begin{figure}[h!]
\centering
\includegraphics[width=0.6\textwidth]{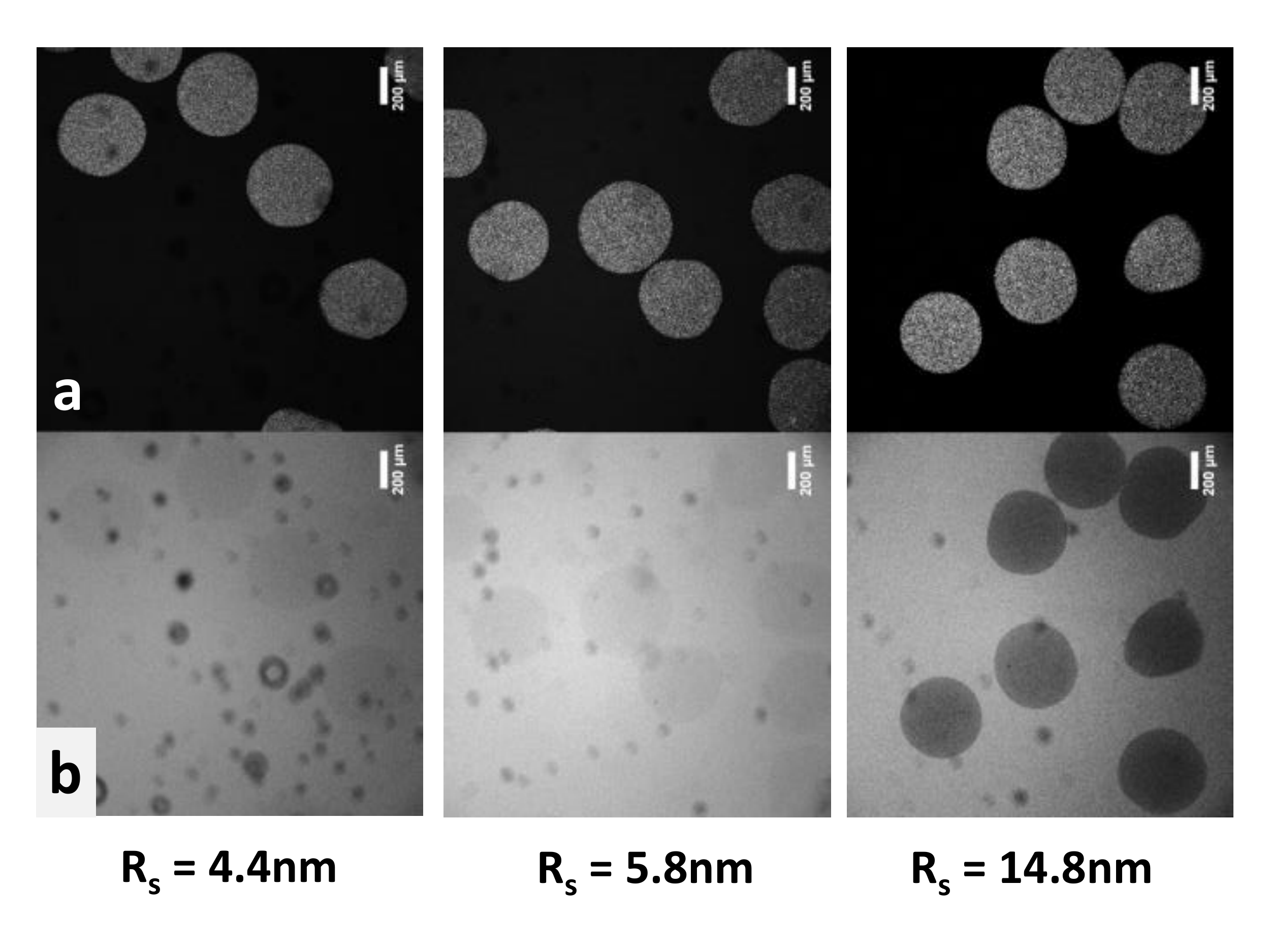}
\caption{\label{fig:MG_Permeability} \textbf{Porosity of MG beads}  (a) MG beads observed by epifluorescence. (b) MG beads dipped in a solution containing fluorescent tracers of different sizes. Small tracers ($R_S$ < 5.8 nm) penetrate the MG beads, while larger tracers ($R_S$=14.8 nm) are excluded. Scale bars = 200$\mu$m.}
\end{figure}

Electron microscopy observations show that pore sizes are extremely heterogeneous in hydrogels. Thus, the typical pore size of MG is difficult to evaluate. However, we can 
empirically define an exclusion-size, above which globular molecules do not penetrate the gel.  To evaluate this exclusion-size, MG beads were prepared according to the protocol in \ref{sec:MG_beads_prepa}. Next, we dipped MG beads in a solution containing fluorescent tracers with different radii. Depending on their size, those tracers either entered the MG or not. In practice, we used fluorescently labelled Dextran tracers with different molecular weights (40, 70 and 500 kDa), corresponding to Stokes' radii ($R_S$) of 4.4 nm, 5.8 nm and 14.8 nm respectively.
We used fluorescent Dextran at a concentration  smaller than 5 $\mu$M. This concentration was sufficient to provide a clearly measurable fluorescence signal, but only exerted a negligible osmotic pressure of ~10 Pa.
Fig.~\ref{fig:MG_Permeability} shows that small tracers with $R_S < 5.8$ nm fully permeate MG beads, as we observe the same level of fluorescence both inside the MG beads and in the surrounding solution. Conversely, beads dipped in a solution containing large tracers ($R_S$ = 14.8 nm) appear darker than the surrounding medium. Large tracers are excluded from the MG. Our results indicate that the MG exclusion-size is in the range between 6 and 14 nm.

\FloatBarrier

\section{\tr{Pore size} of ECM in MCS}\label{sec:poro_ECM}
\begin{figure}[h!]
\centering
\includegraphics[width=0.8\textwidth]{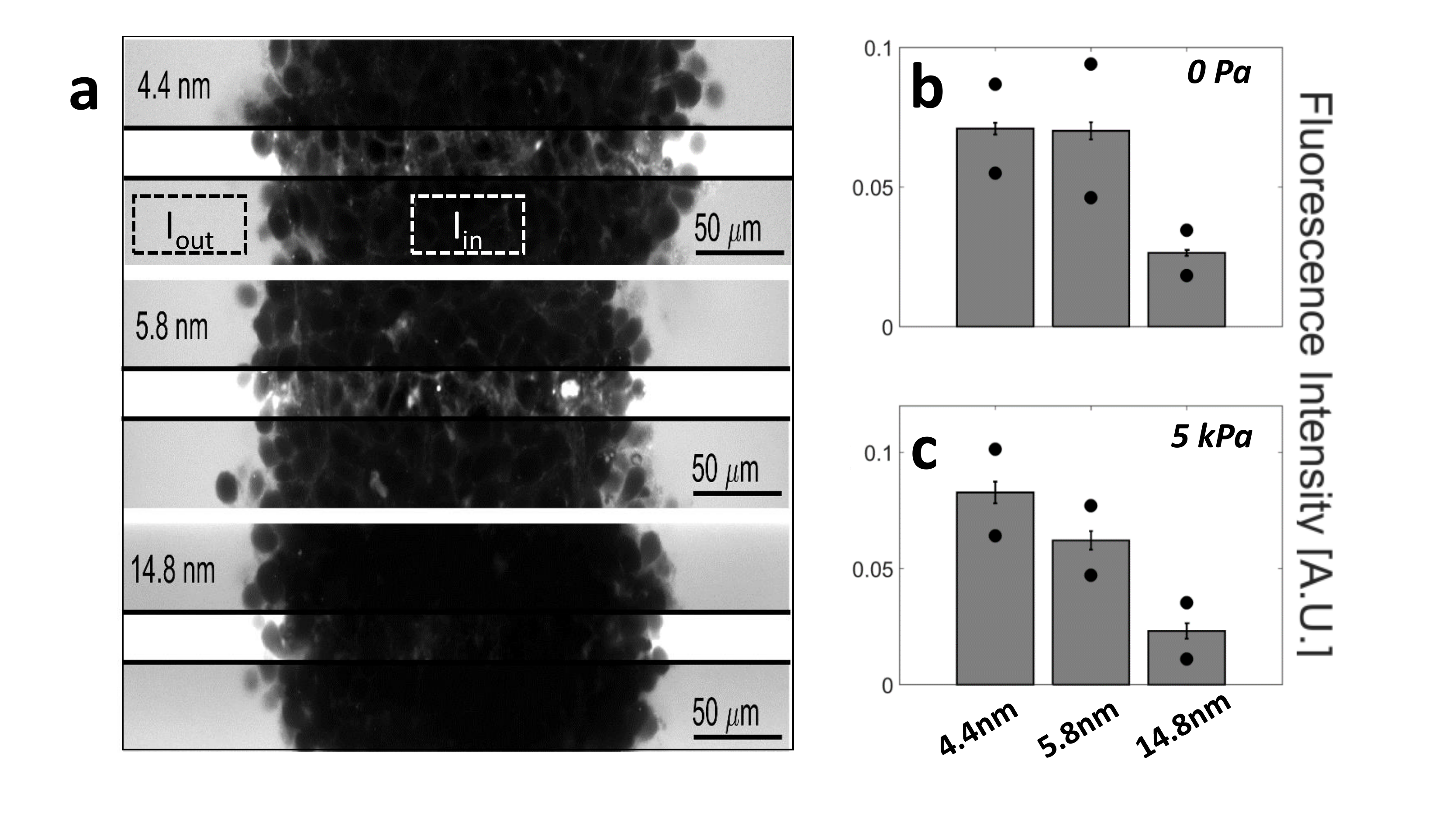}
\caption{\label{fig:MCS_Permeability} \textbf{Exclusion size of the ECM in MCS.} (a) Confocal sections of three MCS dipped in culture media supplemented with Dextran of increasing molecular weights. In order to quantify the total amount of Dextran permeating the MCS, the mean fluorescent intensity measured inside the MCS $\langle I_{In}\rangle$ is normalized by the mean intensity measured in solution $\langle I_{Out}\rangle$. To avoid saturation of $\langle I_{Out}\rangle$, the photomultiplier gain is kept low. This reduces the visibility of extracellular space inside the MCS. In the middle stripe of each image, the brightness is increased of the same amount to make the fluorescence of Dextran visible in the extracellular space. (b) Relative intensity $\langle I_{In}\rangle$/$\langle I_{Out}\rangle$ for different Dextran sizes and without osmotic pressure. Black circles and error bars represent respectively the standard deviation and the standard error of the mean, computed over more than 58 MCS per condition. (c) Relative intensities with an additional osmotic pressure of $\Pi_e=5$ kPa, averaged over 16 (for 4.4 nm), 14 (for 5.8 nm) and 13 (for 14.8 nm) MCS.}
\end{figure}

We observed a similar behavior in MCS dipped in a culture medium supplemented with the same tracers. As shown in Fig.~\ref{fig:MCS_Permeability}~(a), tracers with $R_S$ = 4.4 nm and $R_S$ = 5.8 nm permeated the extracellular space of the MCS but not those larger than 14.8 nm. In order to quantify the relative amount of tracers inside the MCS, we compared the average fluorescence measured inside the MCS $\langle I_{In}\rangle$ and in the surrounding solution $\langle I_{Out}\rangle$. Fig.~\ref{fig:MCS_Permeability}~(b) and Fig.~\ref{fig:MCS_Permeability}~(c) report the relative intensities $\langle I_{In}\rangle$/$\langle I_{Out}\rangle$ , obtained respectively at an external osmotic pressure $\Pi_e = 0$ Pa (N=180) and at $\Pi_e = 5$ kPa (N=43). In both cases, the fluorescence level significantly lowers with large tracers. In terms of pore size, MG is thus a good proxy of ECM, as both have an exclusion size of about 10 nm.
 
\FloatBarrier

\bibliography{spheroid}
\bibliographystyle{model2-names}

\end{document}